\RequirePackage{amsmath}
\RequirePackage{accents} % always after amsmath /!\ conflicts
\documentclass[twocolumn]{svjour3} %smallcondensed referee twocolumn smallextended, envcountsect, final, a4paper, onecolumn, 10pt
\usepackage[round]{natbib} % format biblio , sort&compress

\usepackage[title]{appendix}%to put Appendix prefixes (custom with \appendixname)
\usepackage{placeins} % for floatbarrier
\journalname{Structural Multidisciplinary Optimization}
\usepackage{stmaryrd} % brackets for integer sets [[1,n]]
\usepackage{lmodern} % avoids fonts size warinings
\usepackage{caption}
\usepackage{ulem}
\usepackage{subcaption}
\usepackage[utf8]{inputenc}

\usepackage{graphicx}

\usepackage{amsfonts}
\usepackage{algorithm}
\usepackage[end]{algpseudocode} %noend
\usepackage{xcolor}
\usepackage{tikz}
\usepackage{stanli}
\usepackage{bm}
\usepackage{mathtools}
\usepackage{multirow}
\usepackage{mathtools}
\usepackage{mdframed}

\usepackage{booktabs}
\newcommand{\ra}[1]{\renewcommand{\arraystretch}{#1}}
\usetikzlibrary{external,shapes,arrows,fit}
\makeatletter
\newcommand{\useexternalfile}[2]{%
    \tikzsetnextfilename{#2-output}%
    \scalebox{#1}{\input{\tikzexternal@filenameprefix#2.tex}}}
\makeatother
\newcommand{\ubar}[1]{\underaccent{\bar}{#1}}

\usepackage[colorlinks, linkcolor=black, citecolor=blue]{hyperref}
\usepackage{etoolbox}

\makeatletter
\patchcmd{\NAT@citex}
  {\@citea\NAT@hyper@{%
     \NAT@nmfmt{\NAT@nm}%
     \hyper@natlinkbreak{\NAT@aysep\NAT@spacechar}{\@citeb\@extra@b@citeb}%
     \NAT@date}}
  {\@citea\NAT@nmfmt{\NAT@nm}%
  \NAT@aysep\NAT@spacechar\NAT@hyper@{\NAT@date}}{}{}

\patchcmd{\NAT@citex}
  {\@citea\NAT@hyper@{%
     \NAT@nmfmt{\NAT@nm}%
     \hyper@natlinkbreak{\NAT@spacechar\NAT@@open\if*#1*\else#1\NAT@spacechar\fi}%
      {\@citeb\@extra@b@citeb}%
     \NAT@date}}
  {\@citea\NAT@nmfmt{\NAT@nm}%
  \NAT@spacechar\NAT@@open\if*#1*\else#1\NAT@spacechar\fi\NAT@hyper@{\NAT@date}}
  {}{}
\makeatother
\begin{document}

\title{ An outer approximation bi-level framework for mixed categorical structural optimization problems.}
%\subtitle{Using  the  LaTex Template}
\author{Pierre-Jean Barjhoux
		\and Youssef Diouane 
		\and  Stéphane Grihon
		\and Joseph Morlier}

\institute{Pierre-Jean Barjhoux  \at %\affmark[1]
             Airbus SAS, \\
            Institute of Technology IRT Saint Exup\'ery,\\
             Institut Cl\'ement Ader (ICA)\\
             Toulouse, France \\ \email{pierre-jean.barjhoux@airbus.com}
          \and
          Youssef Diouane  \at
               Department of Mathematics and Industrial Engineering,\\ Polytechnique Montr\'eal.\\
              Mont\'eal, QC, Canada \\ \email{youssef.diouane@polymtl.ca}
          \and
          St\'ephane Grihon  \at
              Airbus SAS \\
              Toulouse, France \\
              \email{stephane.grihon@airbus.com}
          \and
          Joseph Morlier  \at
              Universit\'e de Toulouse, \\
              Institut Cl\'ement Ader (ICA), \\
              CNRS-ISAE SUPAERO-INSA-Mines Albi-UPS \\
              Toulouse, France \\
              \email{joseph.morlier@isae-supaero.fr}
}

\date{Received: date / Accepted: date}
\maketitle
 	\begin{abstract}In this paper, mixed categorical structural optimization problems are investigated. The aim is to minimize the weight of a truss structure with respect to cross-section areas, materials and cross-section type. The proposed methodology consists of using a bi-level decomposition involving two problems: master  and slave. The master problem is formulated as a mixed integer linear problem where the linear constraints are incrementally augmented using outer approximations of the slave problem solution. The slave problem addresses the continuous variables of the optimization problem. The proposed methodology is tested on three different structural optimization test cases with increasing complexity. The comparison to state-of-the-art algorithms emphasizes the efficiency of the proposed methodology in terms of the optimum quality, computation cost, as well as its scalability with respect to the problem dimension. A challenging 120-bar dome truss optimization problem with 90 categorical choices per bar is also tested. The obtained results showed that our method is able to solve efficiently large scale mixed categorical structural optimization problems.	\end{abstract}

%%%%%
%SECTION INTRO
%%%%%
\section{Introduction}\label{sec:introduction_oa}

% problem definition
In this paper, we investigate a class of structural optimization problems with a fixed topology (for the structure) but involving mixed categorical design variables~\citep{Barjhoux2020,Grihon2012, Grihon2017}.
Typically, in the context of structural optimization, the choices of material properties or cross-section types are depicted by categorical variables. The thicknesses or cross-section areas belong to the set of continuous design variables.
Many optimization algorithms are designed to solve such problems.
For example,  metaphor-based metaheuristics and swarm intelligence algorithms \citep{Liao2014, Goldberg1989,Nouaouria2011} natively handle discrete variables. 
However, these methods are not suitable for solving large scale optimization problems \citep{Sigmund2011, Stolpe2011}.

Various surrogate-based optimization strategies have been extended to solve mixed-categorical structural optimization problems \citep{FilomenoCoelho2014,Muller2013,Herrera2014,Roy2017,Roy2019,Garrido-Merchan2018,Pelamatti,Paul_2022,Raul_2022}.
One of the main challenges of such approaches is related to their inefficiency when handling large dimension categorical design space. Other existing works propose new formulations of the original optimization problem by reducing the dimension of a structural optimization problem or by using continuous relaxation of the design variables \citep{Gao2018, Stegmann2005, Krogh2017}. For all these existing approaches, there is no guarantee that the optimization will retrieve the best categorical choices.

By converting categorical variables to integers in the structural optimization problems, classical mixed-integer programming approaches can also be used to solve such problems. In this context, many existing approaches are based on branch-and-bound. For instance,
 \citep{Achtziger2007, Stolpe2007} proposed to rewrite the relaxed problem within their branch-and-bound algorithms as a convex problem which helps to reach easily the global minimum of the relaxed problem.
Other existing approaches are based on decomposition strategies to transform the original problem into a sequence of easy-to-solve subproblems, e.g., Bender decomposition \citep{Benders1962,Geoffrion1972} and outer approximation \citep{Duran1986,Fletcher1994,Hijazi2014}. Several variants based on outer approximation algorithm have been implemented \citep{Stolpe2015} and successfully applied on mixed-integer structural optimization problems.
In the context of continuous structural optimization, decomposition schemes have been widely used, e.g., StiffOpt \citep{samuelides:hal-01422239},  Quasi Separable Decomposition  \citep{Haftka2006,Schutte2004}.

In an industrial context, practical methodologies have emerged to tackle the curse of dimensionality when dealing with categorical variables in large scale structural optimization. For instance, \citep{Grihon2017} uses a bi-step strategy involving massively parallel element-wise optimizations. In fact, by assuming that the main optimization problem is separable with respect to design variables, the approach reduces to a set of optimization problems at the element (subsystem) level, with fixed internal loads. 
This, in particular, simplifies the impact of each categorical choice on the overall optimal internal loads distribution.
This approach is industrially recognized at Airbus. The approach has a computational complexity that depends linearly with respect to the number of structural elements and categorical values. 
Although this existing approach is scalable, it can not handle system-level behavior (optimum internal load distribution) nor system-level constraints (e.g., flutter, modal or displacement constraints). The absence of such constraints in the problem formulation is not representative of aircraft structure design problems, in a multidisciplinary design optimization for instance.

% focus on our work
The proposed methodology in this paper relies on previous works in \citep{Barjhoux2018,Barjhoux2017,Barjhoux2020} where a bi-level methodology was initially proposed. The framework is based on master and slave problems. In \citep{Barjhoux2017}, it has been shown that the hybrid branch-and-bound based approach (for the master problem) can be costly in terms of the number of calls to the finite elements model. 
The exploration cost was shown to grow exponentially with the number of elements and categorical choices, preventing from using this algorithm to solve large scale problem instances. The computational cost of the Bi-level methodology as proposed in \citep{Barjhoux2020} is quasi-linear with respect to the number of structural elements. In particular it permits to solve medium to large scale structural optimization problems (up to two hundred mixed variables).
The latter approach offered an interesting compromise between the quality of the solution and the computational cost, provided the simplicity of the methodology.

In this study, we propose a new Bi-level methodology that leverages the use of linearizations of the subproblems in an Outer Approximation (OA) framework~\citep{Fletcher1994}. This  leads to a more significant computational cost reduction. %further
In our proposed formulation, the mixed categorical-continuous problem is first formulated as a mixed integer-continuous problem with relaxable integer design variables. The continuous design variables are handled by the slave problem while the integer variables are governed by the master problem.
The latter consists of solving a mixed integer linear problem built iteratively by concatenating linear approximations of the slave problem solutions.
This approach is different from the OA framework originally presented in \citep{Fletcher1994} in the sense that only a linear approximation of the slave problem is used to define the master problem approximation.
The derivatives used to build the linear approximations of our subproblems are constructed using post-optimal sensitivities~\citep{Fiacco1976}. Under a convexity assumption, we will show that the proposed approach converges to the optimal solution. Although, the convexity assumptions cannot be verified for structural optimization problems in general, the obtained numerical results show that our proposed method is performing very well compared to state-of-the-art methods.

% plan
This paper is organized as follows. In Section \ref{sec:problem_statement_OA}, the formulation of the mixed categorical-continuous optimization problem is presented.
In Section \ref{sec:methodology_oa}, the proposed methodology is presented.
The performance and the scalability of our approach are compared with state-of-the-art algorithms in Section \ref{sec:numerical_results_oa}. The section is concluded with the obtained results on a structural problem of 120 structural elements with 90 categorical choices for each of element. Conclusion and perspectives are drawn in Section~\ref{sec:conclusion_oa}.

%%%%
% SECTION PROBLEM DEFINITION
%%%%
\section{Problem statement}
\label{sec:problem_statement_OA}

\subsection{Design space}
\label{ssec:design_space}

In this work, our goal is to minimize the weight of a structure, at fixed topology, by exploring the internal geometry as well as material description of all the structural elements of the problem. 
Two kinds of design variables are thus involved when handling these kind of problems.

Firstly, categorical choices are involved as design variables during the optimization process. 
Indeed, in this work, the possible choices of material and member profiles (e.g., ``I'', ``C'', ``T'') for each element will be regarded as a part of the design variables and that have to be explored. 
The categorical choices will be described by a matrix $\vec B$ that has $n \times p$ binary coefficients, where $p$ is the possible number of choices per structural element $n$. In this paper, we will often refer to the set possible choices by the set of catalogs. We specify here that the $p$ available catalogs are the same for all the structural elements.
In this context, during the optimization we assign to each element a choice of material and cross-section type, all described by one categorical design variable. 
Let $\mathcal C^{n\times p}$ be the enumerated set that contains the $p^n$ choices of materials and member profiles for each element of the structure, so that :
\begin{eqnarray*}
%\begin{aligned}
\mathcal C^{n\times p} &:=& \left \{ X \in \mathbb{R}^{n \times p} ~:~ X_{ij}  \in \{0,1\}~\mbox{and}~ \sum_{j=1}^{p} X_{ij} = 1 \right\},
%\end{aligned}
\end{eqnarray*}
where $X_{ij}$ represents the element at the $i^{\mbox{th}}$ row and $j^{\mbox{th}}$ column of the matrix $X$. 
$X_{ij}$ is a binary choice variable among the existing $p$ choices per structural element, with $X_{ij}$ = 1 if for the $i^{th}$ element the $j^{th}$ categorical option is chosen and $X_{ij}$ = 0 otherwise.
We note that,  each row of the matrix $X \in \mathcal C^{n\times p}$ describes the catalog choices composition of a given element.
For example, if $\vec B \in \mathcal C^{10\times 4}$ and $ B_{42} = 1$, then the categorical choice associated to the $4^{th}$ structural element is the choice $2$, corresponding to a given combination of profile and material among the $4$ available choices.

\begin{figure}
\centering
 \includegraphics[scale=.23]{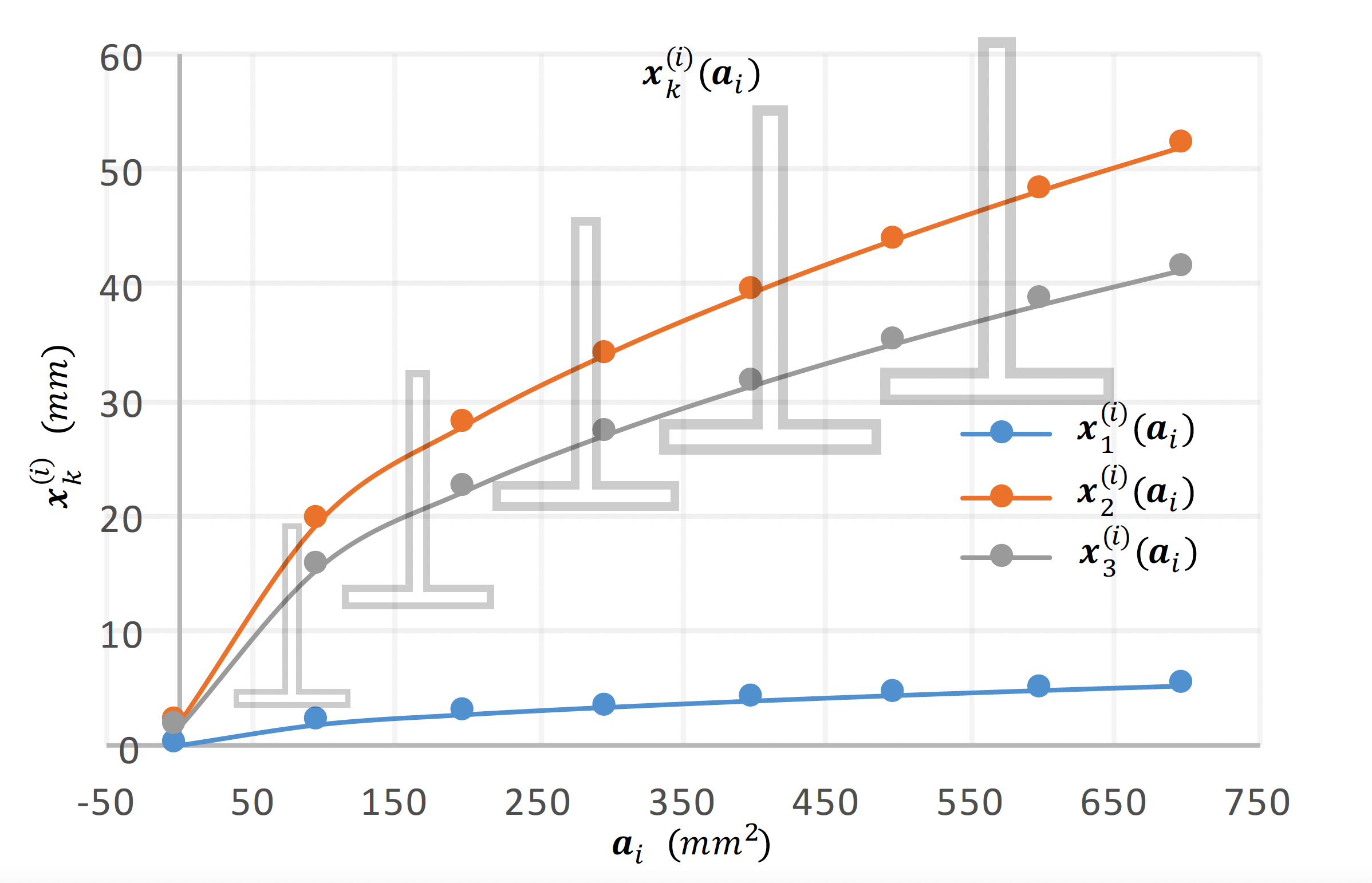}
\caption{Scaling of a bar section. Example with ``T''-profile.}
\label{fig:detailed_geometry_scaling}
\end{figure}

In a second time we treat the member profiles areas as continuous design variables. Formally, the areas can be represented as a vector $\vec{a} \in \mathbb R^n$ where the number of components $n$ corresponds to the number of structural elements. For a given choice of member profile, the areas scale the internal shape of the structural elements \citep{Barjhoux2017, Barjhoux2018}.
The description of the internal cross-section parameters (with respect to areas $\vec a_i$) is given by
\begin{align}\label{eq:detailed_variables}
	\vec x^{(i)}(\vec a_i) := \sqrt{\frac{\vec a_i}{\vec a_0(\vec B_{i,:})}} \vec x_0(\vec B_{i,:}),
\end{align}
where $\vec{B}_{i,:}$ represents the $i^{th}$ row of the matrix $\vec{B}$ and $\vec x_0 (\vec B_{i,:})$ is the reference detailed geometry of the profile driven by the choice $\vec B_{i,:}$. This way, the  parameters $\vec x^{(i)}$ are estimated as latent variables that depend on the areas $\vec a_i$.
This description of the internal member profile geometry is inspired from existing approaches like for example the PRESTO methodology in \citep{Grihon2012, Gao2018}.
Fig. \ref{fig:detailed_geometry_scaling} shows how internal parameters (and so the area moments of inertia) can be scaled using the area of the cross-section. The proposed scaling allows to handle any profile type as far as we can add new member profiles depending on the design space we want to explore.

\subsection{Objective and constraints functions}
\label{ssec:objective_constraints_functions}
The objective function and the constraints as presented in \citep{Barjhoux2020} are reformulated in this article in particular for the sake of clarity.
First, the categorical variable is coded as a binary variable.
Second, we use continuous definitions of objective and constraints functions. 
The binary variables are introduced as continuous weighting factors in these functions. This means that, each of the objective and constraints functions can be evaluated at intermediate values of $\vec B$, even if the outputs have no physical meaning.

First, we need to define the space on which the functions are defined.
Let $\widetilde{\mathcal{C}}^{n \times p}$ be the set of matrices $\vec B$ of real coefficients such as :
\begin{eqnarray*}
%\begin{aligned}
\widetilde{\mathcal{C}}^{n \times p} &:=& \left \{ X \in \mathbb{R}^{n \times p} ~:~ X_{ij}  \in [0,1]~\mbox{and}~ \sum_{j=1}^{p} X_{ij} = 1 \right\}.
%\end{aligned}
\end{eqnarray*}
In other terms, $\widetilde{\mathcal{C}}^{n \times p}$ is the continuous relaxation of $\mathcal{C}^{n \times p}$ on $[0,1]$.
Of course, there is no underlying physical meaning when $\vec B$ takes intermediate (real) values in $[0,1]$.
The values of $\vec B$ in $\widetilde{\mathcal{C}}^{n \times p}$ will serve as weighting factors in the objective and constraints functions.

In this problem, the objective is the weight function, given by :
\begin{eqnarray}
\widetilde{w} \colon & \mathbb{R}^{n} \times \widetilde{\mathcal{C}}^{n \times p} &\to \mathbb{R} \nonumber \\
& (\vec a, \vec B) &\mapsto \sum_{i=1}^{n} \sum_{c=1}^{p}  \rho(c) \ell_i {\vec B}_{ic}  \vec a_i,
\label{eq:weight_function}
\end{eqnarray}
with $\rho(c)$ refers the density of the material that corresponds to the choice $c$. 
It is computed as the sum of the $p$ available densities weighted by the continuous choices~$\vec B$. the constant $\ell_i$ denotes the length of element $i$.

The constraints $\bm \delta$ on displacements $\vec{u}$
ensure that on $d$ given nodes of the truss the displacements will not exceed predefined upper bounds $\vec{\bar{u}} \in \mathbb{R}^d$. 
With $\bm P$ a projector that select the elements on which the displacement constraint will apply, the definition of $\bm \delta$ function is given as follows :
\begin{eqnarray}
  \widetilde{\bm \delta} \colon & \mathbb{R}^{n} \times \widetilde{\mathcal{C}}^{n \times p} &\to \mathbb{R}^d \nonumber \\
 & (\vec{a}, \vec{B}) &\mapsto \bm P \bm u(\vec{a}, \vec{B}) - \bar{\vec{u}}. %\widetilde{\vec E}
\end{eqnarray}

The stress constraints $\widetilde{\vec{s}}_{ij}$ are defined such as :
\begin{eqnarray}
\widetilde{\vec{s}}_{ij} \colon &\mathbb{R}^{n} \times \widetilde{\mathcal{C}}^{n \times p} &\to \mathbb{R} \nonumber \\
&(\vec a, \vec B) &\mapsto \sum_{c=1}^p \vec{B}_{ic} \vec s_{ij}(\vec{a}_i, c,
\vec{\Phi}_i(\vec{a}, \vec{B})) %\widetilde{\vec E}
\end{eqnarray}
where $\vec s_{ij}$ is given by :%\begin{equation}
\begin{align*}
  \bm s \colon \mathbb{R}^{n} \times \widetilde{\mathcal{C}}^{n \times p} &\to \mathbb{R}^{n\times m}
\end{align*}
and is of the form, for every choice $c$ among the categorical set $\{1, \dots, p\}$
$$
\begin{pmatrix}
\bm s_{11}(\vec{a}_1, c, \vec{\Phi}_1(\vec{a},\vec{B})) & & \dots  & & \bm s_{1m}(\vec{a}_1, c, \vec{\Phi}_1(\vec{a},\vec{B})) \\
\bm s_{21}(\vec{a}_2, c, \vec{\Phi}_2(\vec{a},\vec{B}))& & \dots & & \bm s_{2m}(\vec{a}_2, c, \vec{\Phi}_2(\vec{a},\vec{B})) \\
 \vdots & & \ddots & & \vdots \\
 \bm s_{n1}(\vec{a}_n, c, \vec{\Phi}_n(\vec{a},\vec{B})) & & \dots & &\bm s_{nm}(\vec{a}_n, c, \vec{\Phi}_n(\vec{a},\vec{B})) 
\end{pmatrix}.
$$
%with $\vec \Phi_i$ the internal forces on the element $i$ named ${elt_i}$. 
The element constraint $\bm s_{ij}$ is defined as the difference between the structural member stress constraints value and a limit stress. In particular, if the members constraint stress exceed the limit stress, then  $\bm s_{ij}$ will take negative values and the constraints will be violated. Practical expressions of these optimization constraints are provided in the numerical section (see \eqref{allowable_constraints1}, \eqref{allowable_constraints2}, \eqref{buckling_constraints1} and \eqref{buckling_constraints2}).

In the context of this work, there is no change in the topology of the structure. Internal forces $\vec \Phi$ and displacements $\vec u$ will be computed using the direct stiffness method, introduced in \citep{Turner1959,Turner1964}. Structural elements are considered as truss elements with pin-jointed connections. This means that the bars will only carry axial forces. The cross-section shapes will be only involved through the Euler and local constraints definition (see Section~\ref{sec:numerical_results_oa}). The stiffness matrix of the structure is not impacted by the cross-section shapes.
At each node, displacements are allowed along the global axes. Each element $i$ is defined by the elementary stiffness matrix $\bm K^e_i(\vec{a}_i,\vec{B}_{i,:}) \in \mathbb R^{q,q}$, with $q$ the number of free nodes multiplied by the number of physical space dimensions.
The global stiffness of the whole truss is given by the matrix $\bm K (\vec{a},\vec{B}) \in \mathbb R^{q,q}$ in global coordinates. Such matrix can be computed as the sum of each element stiffness matrix expressed after its transformation with the $i^{th}$ element rotation matrix $\vec T_i$, i.e., \citep{Turner1959,Turner1964}:
\begin{align*}
    \bm K (\vec{a}, \vec{B}) := \sum_{i=1}^{n} [ \vec T_i^{\top} \bm K^e_i(\vec{a}_i,\vec{B}_{i,:}) \vec T_i ] .
\end{align*}
Given a vector $\bm f \in \mathbb{R}^{q}$ of external loads applied on each of the free nodes in the global coordinates, the vector of displacements $\bm u \in \mathbb{R}^{q}$  can be obtained by solving the following equation:
\begin{align}
    \label{eq:forces_equilibrium}
    \bm K(\vec{a},\vec{B})  \bm u(\vec{a},\vec{B}) = \bm f.
\end{align}
The vector of internal forces $\vec \Phi \in \mathbb{R}^{n}$ is then given by:
\\
\noindent $\forall i \in \{1, \dots, n\}$,
\noindent \begin{align*}
    & \vec{\Phi}_i(\vec{a},\vec{B}) := \bm K^e_i(\vec{a}_i,\vec{B}_{i,:}) \vec T_i \bm u_i(\vec{a},\vec{B}),
\end{align*}
where $\vec \Phi_i$ is the axial force through element $i$ and $\bm u_i$ its displacement vector.

\subsection{Optimization problem}
\label{ssec:optimization_problem}

As formulated in \citep{Barjhoux2020}, the problem involves categorical non-ordered and non-relaxable design variables.
The formulation prevents from using algorithms that exploit the gradient of the functions with respect to all the design variables.
This is why, objective and constraints have been introduced as continuous functions in Section \ref{ssec:objective_constraints_functions}.
The categorical optimization problem is a mixed categorical continuous optimization problem, formulated as a mixed integer non linear programming (MINLP) problem.
The optimization problem consists of a structural weight minimization with respect to stress and displacements constraints :
\begin{equation*}\tag{P}
\label{pb:OverallPbBinary}
\begin{aligned}
& \underset{(\vec a, \vec B) \in  \mathbb{R}^n \times \mathcal{C}^{n \times p} }{\text{minimize}} & & \widetilde{w}(\vec a,\vec B) \\ 
& \underset{}{\text{subject to}}
& & \widetilde{\vec s}(\vec a,\vec B) \leq \vec 0_{n,m}  \\
& & & \widetilde{\vec \delta}(\vec a,\vec B) \leq \vec 0_{d}\\
& & & \ubar{\vec a} \leq \vec a \leq \bar{\vec a} \\
%& & & \vec B_{ij} (\vec B_{ij} - 1) = 0 
\end{aligned}
\end{equation*}
where $\ubar{\vec a} \in \mathbb{R}^n$ and $\bar{\vec a} \in \mathbb{R}^n$ are the lower and upper bounds on areas, respectively.
It is worth to note that the design space is $\mathcal{C}^{n \times p}$, such that the solution $\vec B$ is binary.
The methodology presented in the Section \ref{sec:methodology_oa} will take advantage of the continuous property of the objective and constraints functions. In the next section, we will describe our proposed methodology to solve \eqref{pb:OverallPbBinary}.

%%%%%%
%
% SECTION METHODOLOGY
%
%%%%%%

\section{Methodology}\label{sec:methodology_oa}

In what comes next, for the sake of clarity, we will consider $\vec B$ and $\widetilde{\vec s}$ as vectors instead of matrices. Namely, $\mathcal{C}^{n \times p}\subset  \mathbb{R}^{np}$ and $\mathbb{R}^{n \times m} \sim \mathbb{R}^{nm}$. The identification between the vector and matrix space can be trivially obtained, for instance,  for any $\vec A \in \mathbb{R}^{n \times p}$ one can use
\begin{align*}
[\vec A_{11}, \dots, \vec A_{1p}, \vec A_{21}, \dots, \vec A_{2p}, \dots, \vec A_{n1}, \dots, \vec A_{np}]^{\top}
\end{align*}
to consider it as an element of $\mathbb{R}^{np}$ as well. We note  that in our optimization setting the topology will be kept unchanged. Handling the change in the topology of the structure will not be covered by our approach. The reason behind this restriction will be explained in Section~\ref{sssec:gradient_psi_oa}.

\subsection{A bi-level framework} \label{ssec:decomposition_oa}

The proposed bi-level decomposition of the problem (\ref{pb:OverallPbBinary}) is presented in this section.

For a given $\vec B \in \mathcal{C}^{n\times p}$, let $\Omega (\vec B) \subset \mathbb{R}^n$ be the set of feasible constraints of the problem (\ref{pb:OverallPbBinary}) given by 
\begin{equation*}
\label{eq:omega_binary}
\begin{aligned}
\Omega(\vec B):=\{\ubar{\vec{a}} \leq \vec{a} \leq \bar{\vec{a}}~:~\widetilde{\bm s}(\vec{a},\vec{B}) \leq \bm 0_{mn}~\mbox{and}~\widetilde{\bm \delta}(\vec{a},\vec{B}) \leq \bm 0_{d}
\}.
%& \vec{B}_{ij} = 0 \text{ or } 1
\end{aligned}
\end{equation*}
An efficient way to solve pure continuous optimization problems is by taking advantage of gradient based algorithms.
In the problem introduced in Section \ref{sec:problem_statement_OA}, it can be seen that by fixing (temporarily) design variables $\vec B$ in (\ref{pb:OverallPbBinary}) at integer values, the optimization problem becomes a continuous one parameterized with $\vec B$, and where integrity constraints on $\vec B$ can be removed. 
This means that at a given $\vec{B}$, the weight $\widetilde{w}$ can be minimized with respect to the remaining continuous design variables that are the areas $\vec a \in \Omega(\vec B)$. 
This leads to the following slave problem (\ref{pb:binary_parameterized_continuous_opt}), that reduces to a structural sizing optimization problem:
\begin{align*}\tag{sP($\vec{B}$)}
\Psi(\vec{B})~:=~\underset{\vec{a} \in \Omega(\vec B)}{\text{minimize}}~\widetilde{w}(\vec a, \vec B).
\label{pb:binary_parameterized_continuous_opt}
\end{align*}
The structure of the problem is such that this remaining optimization problem
becomes more tractable. In fact, the decomposition leverages the use of the gradients (with respect to $\vec a$) of the objective and constraints to solve the problem (\ref{pb:binary_parameterized_continuous_opt}).
This is the main motivation in handling the continuous variables separately from the integer ones.
In this approach, the integer (binary) variables will be handled by a master problem (\ref{pb:binary_bendersDecomposition_up}) of the form 
\begin{align*} \tag{mP}
\label{pb:binary_bendersDecomposition_up}
\underset{\vec B \in \mathcal C^{n \times p}}{\text{\textup{ minimize}}} ~~ & \Psi(\vec B), 
\end{align*}
with $\Psi(\vec{B})$ is the result of the slave Problem (\ref{pb:binary_parameterized_continuous_opt}).
The slave problem (\ref{pb:binary_parameterized_continuous_opt}) takes these complicating variables $\vec B$ as parameters while optimizing with respect to continuous design variables. 
This means that during the slave optimization, the choices of materials and cross-section types for all elements remain fixed.
This slave problem will be solved using a gradient based method. 
The obtained solution can be seen as a function $\Psi(\vec{B})$ which is parameterized by the categorical choices through the continuous coding $\vec B$. 
Namely, $\Psi(\vec{B})$ corresponds to the optimal weight
of the slave problem knowing the variables $\vec{B}$. 
This function is then taken as the objective of the master optimization problem (\ref{pb:binary_parameterized_continuous_opt}).
Although the slave problem can be easy to handle using gradient-based algorithms, the difficult part remains in the master problem.
In fact, the problem (\ref{pb:binary_bendersDecomposition_up}) is still a large-scale pure integer non-linear optimization problem, that usual combinatorial optimization solvers fail to solve efficiently.
However, unlike the problem presented in \citep{Barjhoux2020}, the integer variable $\vec B$ is relaxable and the functions are defined at intermediate non 0-1 values of $\vec B$.
Moreover, all the functions of the optimization problem are continuously differentiable.
This is a basic requirement to compute the sensitivity of the slave problem solution parameterized in~$\vec{B}$.

\subsection{On the minimization of $\Psi$} \label{ssec:bilevel_outer_approx_algo}

In this paper, we suggest to solve the master problem (\ref{pb:binary_bendersDecomposition_up}) by means of outer approximation (OA) cuts that are built using the gradient information on $\Psi$.
We propose to consider at the master level the minimization of an approximated problem $\mathcal P$ instead of (\ref{pb:binary_bendersDecomposition_up}), so that the computational complexity of the master problem can be significantly reduced.
For that, the following iterative scheme is implemented. Given an iteration $k$, the master problem (\ref{pb:binary_bendersDecomposition_up}) of the bi-level formulation is reduced to a problem $\mathcal P^{(k)}$ easier to solve.

The slave optimization problem is defined by fixing the binary variables $\vec B^{(k)}$ in the problem (\ref{pb:binary_bendersDecomposition_up}). Due to the set of constraints, the problem (\ref{pb:binary_bendersDecomposition_up}) can be seen as a full integer optimization problem. 
The slave problem reduces to an evaluation of the objective $\Psi(\vec B^{(k)})$ which represents the optimal weight solution of (\ref{pb:binary_parameterized_continuous_opt}) given for a fixed categorical choice $\vec B^{(k)}$. Assuming that there is at least one feasible solution depending on the fixed point $\vec B^{(k)}$, the optimal objective value of the slave problem is an upper bound of the solution to (\ref{pb:OverallPbBinary}). 

A definition of the master problem is given as follows. In fact, under the assumption that the function $\Psi$ is convex, using \cite[Theorem 1]{Fletcher1994}, one deduces that solving the problem (\ref{pb:binary_bendersDecomposition_up}) is equivalent to solving the following mixed integer linear program (MILP) given by 
{\small
\begin{equation}%\tag{OAP($K^*$)}
\begin{aligned}
& \underset{\vec B \in \mathcal C^{n \times p}, \eta \in \mathbb{R}}{\text{\textup{ minimize}}} && \eta \\ 
& \underset{}{\text{~~~~~\textup{s. t.}}} && \eta \geq \Psi(\widetilde{\vec B}) + \dfrac{d\Psi}{d\vec B}\Bigg|_{\widetilde{\vec B}}^{\top}~(\vec B - \widetilde{\vec B}), ~~\forall \widetilde{\vec B} \in C^{n \times p}
\end{aligned}
\label{pb:outer_approx_k}
\end{equation}
}
We note that in the problem (\ref{pb:outer_approx_k}), the function $\Psi$ is replaced by an hyperplane that is also its linear support at $\widetilde{\vec B} \in C^{n \times p}$.

Solving the MILP problem (\ref{pb:outer_approx_k}) directly may be out of reach as it would require $p^n$ evaluations of $\Psi$ corresponding to all integer vectors $\widetilde{\vec B} \in C^{n \times p}$. in our case it would require evaluations of the sizing problem (\ref{pb:binary_parameterized_continuous_opt}) taken at every $p^n$ combinations of materials and cross-section types available in $\{1, \dots, p \}^n$. The number of constraints $p^n$ related to the problem (\ref{pb:outer_approx_k}) can be also extremely large for reasonable values of $p$ and $n$.
For this reason, instead of considering the problem (\ref{pb:outer_approx_k}), the OA algorithm involves a sequence of less expensive relaxed variant of the MILP problem (\ref{pb:outer_approx_k}), i.e., for a given iteration~$(k)$ one solves

{\small
\begin{equation}%\tag{OAP($K^*$)}
\begin{aligned}
& \underset{\vec B \in \mathcal C^{n \times p}, \eta \in \mathbb{R}}{\text{\textup{ minimize}}} && \eta \\ 
& \underset{}{\text{~~~~~\textup{s. t.}}} && \eta \geq \Psi(\widetilde{\vec B}) + \dfrac{d\Psi}{d\vec B}\Bigg|_{\widetilde{\vec B}}^{\top}~(\vec B - \widetilde{\vec B}), ~~\forall \widetilde{\vec B} \in K^{(k)}\\
%& & & \vec B \circ (\vec B - \vec 1) = \vec 0_{np}.
\end{aligned}
\label{pb:relaxed_outer_approx_k_noUBD}
\end{equation}
}
with $K^{(k)}$ a set of $k$ elements in $\mathcal C^{n \times p}$, such that $
    K^{(k)} \subset \mathcal C^{n \times p}.$ We note that, under the convexity assumption of $\Psi$, the problem (\ref{pb:relaxed_outer_approx_k_noUBD}) yields a lower bound to the solution of the Problem (\ref{pb:outer_approx_k}).
At each iteration of the OA algorithm, the problem (\ref{pb:relaxed_outer_approx_k_noUBD}) can be geometrically interpreted as an exploration of the effects of the outer approximations (i.e., the linear supports) on the objective $\Psi$. The set $K^{(k)}$ will be updated recursively as follows  $K^{(k)} \gets K^{(k-1)}~\bigcup~\{ \vec B^{(k)} \}$ where $\vec B^{(k)}$ is the solution of the MILP problem \eqref{pb:relaxed_outer_approx_k_noUBD} at a fixed iteration~$(k)$. Hence, the MILP problem \eqref{pb:relaxed_outer_approx_k_noUBD}  for the iteration~$(k)$ is obtained just by adding to the problem \eqref{pb:relaxed_outer_approx_k_noUBD} (related with the iteration $(k-1)$) the linear  constraint
\begin{equation} \label{constraints:iter:k}
\eta \geq \Psi( \vec B^{(k)} ) + \dfrac{d\Psi}{d\vec B}\Bigg|_{ \vec B^{(k)}}^{\top}~(\vec B -  \vec B^{(k)}).
\end{equation}
A key ingredient for setting the latter constraint is the estimation of the derivative $\frac{d\Psi}{d\vec B}\Big|_{\vec B^{(k)}}$. Under reasonable assumptions, the next subsection details how post-optimal sensitivities can be useful on estimating the gradient of $\Psi$ at $\vec B^{(k)}$.
\subsubsection{Computing the gradient of $\Psi$ at $\vec B^{(k)}$:  $\frac{d\Psi}{d\vec B}\Big|_{\vec B^{(k)}}$}\label{sssec:gradient_psi_oa}

 The estimation of the gradient of $\Psi$ with respect to the parameters $\vec B$ is a key ingredient in setting the constraint \eqref{constraints:iter:k}. 
On that way, the estimated gradient will provide information on the behavior of the optimal weight, solution of (\ref{pb:binary_parameterized_continuous_opt}), subject to a small perturbation of $\vec B^{(k)}$. In this case,
the gradient of $\Psi$ is known as \textit{post-optimal sensitivity} \citep{Fiacco1976}.
It can be noted that such perturbation has no physical meaning since $\vec B$ describes categorical choices in a continuous manner.
The efficient computation of the gradient of $\Psi$ at a given $\vec B^{(k)}$ will be a key feature of our proposed methodology.
Indeed, if the gradient is estimated by finite differences, its computational cost is growing proportionally to $np$ (i.e., the number of parameters $\vec B$). This would typically require to solve $n(p-1)$ optimization problems instances. Hence, for large scale optimization problems, using finite differences may be out of reach.
In the context of large scale optimization problems, estimating the gradient using  post-optimal sensitivity analysis can be very helpful.  In Appendix \ref{apdx:postop_sensitivities},
we give the details of deriving the derivatives $\frac{d\Psi}{d\vec B}\Big|_{\vec B^{(k)}}$.

The post-optimal sensitivity analysis is derived using the Karush-Kuhn-Tucker (KKT) conditions where, particularly, we require the constraint qualification (i.e., the gradients of the constraint functions of active constraints being linearly independent). Such assumption is reasonable in the context of structural optimization problems with a fixed topology. But, it might not be guaranteed for problems where a change in topology is allowed; mainly due to the presence of the so-called vanishing constraints. For this reason, in this work, we consider that the topology of the structure is unchanged during the optimization process.

\subsubsection{An outer approximation bi-level framework}\label{ssec:oa_algo}
\begin{figure*}
\begin{center}
%\useexternalfile{1}{workflow_full}
 \includegraphics[scale=.7]{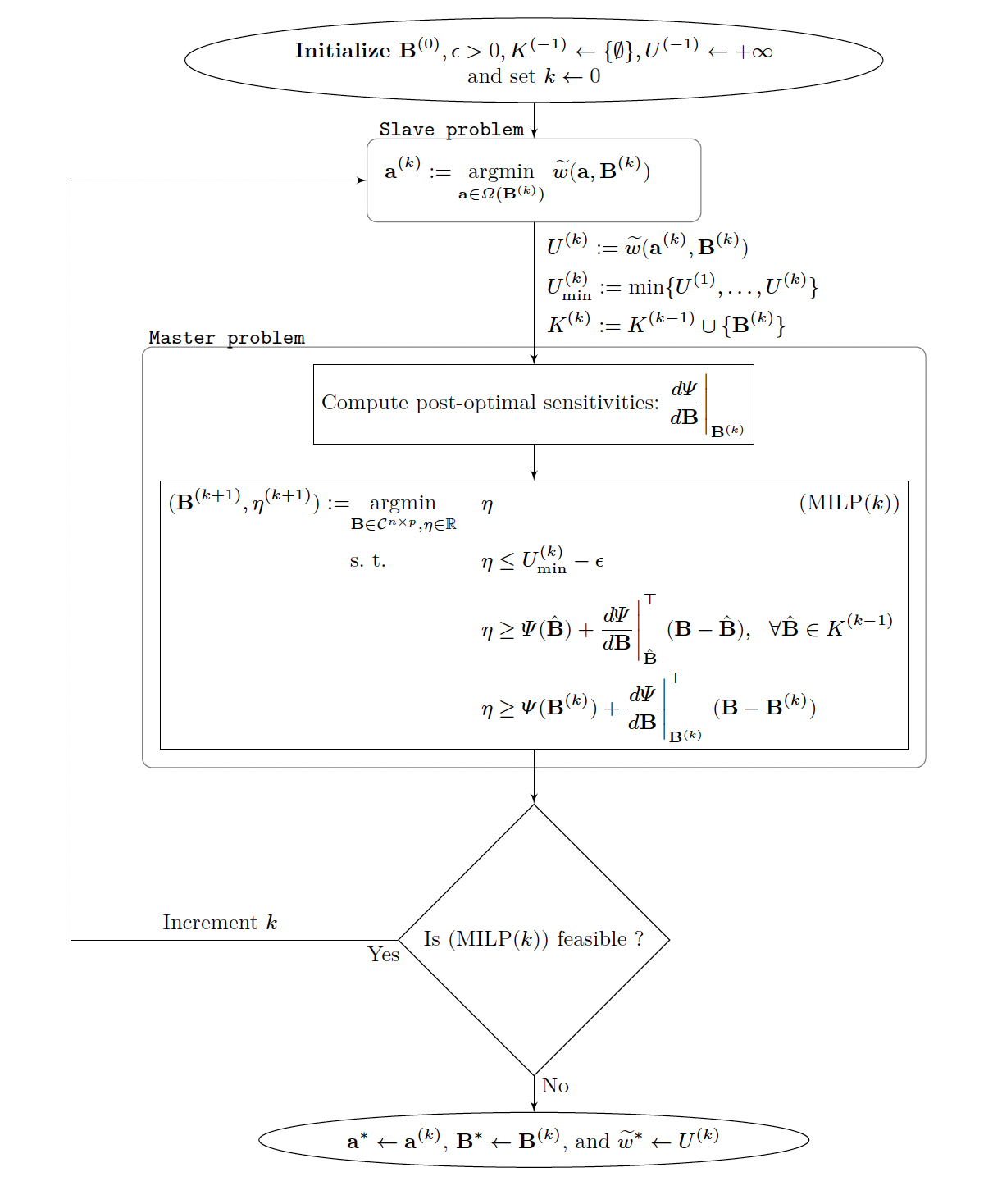}
\end{center}
\caption{Illustration of the proposed methodology (see Algorithm \ref{alg:bilevel_oa}).}
\label{fig:workflow}
\end{figure*}

The proposed algorithm consists of solving an alternating sequence of slave and master problems, as defined previously.
The post-optimal sensitivities of the slave problem (sizing) are involved in the definition of the master problem.
Let $(k)$ be the current outer iteration of the algorithm.
The algorithm workflow is illustrated in Fig. \ref{fig:workflow}. 

First, the slave problem, reduced to an evaluation of $\Psi$, is solved at $\vec B^{(k)}$. 
This means that as a first step, the slave continuous optimization problem (\ref{pb:binary_parameterized_continuous_opt}) aiming at minimizing the weight while satisfying stress and displacements constraints is solved.
The problem (\hyperref[pb:binary_parameterized_continuous_opt]{sP($\vec B^{(k)}$)}) is solved and yields a solution $\vec a^{(k)}$ such that :
\begin{align}
  \vec a^{(k)} := \underset{ \vec{a} \in \Omega(\vec B^{(k)}) } {\text{argmin}}~\widetilde{w}(\vec a, \vec B^{(k)}).
\end{align}
An upper bound $U^{(k)}$ to the solution of (\ref{pb:OverallPbBinary}) is defined by :
\begin{align*}
    U^{(k)} := \widetilde w (\vec a^{(k)}, \vec B^{(k)})= \Psi(\vec B^{(k)}),
\end{align*}
The best current solution of the original problem (\ref{pb:OverallPbBinary}) is thus given by the best upper bound returned during the ($k$) outer iterations :
\begin{align}
    U^{(k)}_{\min} := \text{min} \{U^{(1)}, \dots, U^{(k)}\}.
\end{align}

Second, the relaxed MILP problem (\ref{pb:relaxed_outer_approx_k_noUBD}) can be set up. 
Its definition relies on the linearizations of $\Psi$ taken at the solutions yielded during the ($k$) previous iterations.
While the linearizations from the previous iterates $(l) < (k)$ remain unchanged, the linearization of $\Psi$ at the current iteration $(k)$ has to be computed. 
More precisely, the gradient of $\Psi$ taken at $\vec B^{(k)}$ has to be evaluated.

Once the linearization of $\Psi$ has been computed, it is added as constraint in the problem (\ref{pb:relaxed_outer_approx_k_noUBD}).
Furthermore, since in practice the problem (\ref{pb:OverallPbBinary}) does not need to be solved exactly, it is sufficient to generate the new ($\vec B^{(k+1)}$) by adding a tolerance $\epsilon$ on the upper bound $ U^{(k)}_{\min}$ as an additional constraint to the MILP master problem.
The resulting mixed integer linear integer problem (\ref{pb:relaxed_outer_approx_k}), is thus given by:
{\small{
\begin{equation*}\tag{MILP($k$)}
\begin{aligned}
& \underset{\vec B \in \mathcal C^{n \times p},\eta \in \mathbb{R}}{\text{minimize}} && \eta \\ 
& \underset{}{\text{\textup{s. t.}}} && \eta \leq   U^{(k)}_{\min} - \epsilon \\
& && \eta \geq \Psi(\hat{\vec B}) + \dfrac{d\Psi}{d\vec B}\Bigg|_{\hat{\vec B}}^{\top}~(\vec B - \hat{\vec B}), ~~\forall \hat{\vec B} \in K^{(k-1)}\\
& && \eta \geq \Psi(\vec B^{(k)}) + \dfrac{d\Psi}{d\vec B}\Bigg|_{\vec B^{(k)}}^{\top}~(\vec B - \vec B^{(k)})
\end{aligned}
\label{pb:relaxed_outer_approx_k}
\end{equation*}}}
with $K^{(k)}$ such that
\begin{align*}
    K^{(k)} := K^{(k-1)} \cup \{ \vec B^{(k)} \},
\end{align*} 
and $K^{(k-1)}$ the set of the $k-1$ previous $\vec B^{(k-1)}$.
The problem (\ref{pb:relaxed_outer_approx_k}) is built iteration per iteration by adding, as constraints, linearization of the functions $\Psi$ taken at the current solution $(\vec B^{(k)})$. 
The optimality of the algorithm relies on the convexity of $\Psi$, ensuring that the linearizations are underestimators of $\Psi$. 
Once built, the problem (\ref{pb:relaxed_outer_approx_k}) is solved and provides a lower bound of (\ref{pb:OverallPbBinary}).  Iteratively, the number of constraints within the problem (\ref{pb:relaxed_outer_approx_k}) is getting higher. This ensures a monotonic increase in the lower bound over the iterations (i.e., $\eta^{(k)} \le \eta^{(k+1)}$).

The algorithm will be declared as convergent when the feasible domain of the problem (\ref{pb:relaxed_outer_approx_k}) is getting empty. This particularly means that
the numerical solutions $\vec B^{(k+1)}$ and $\eta^{(k+1)}$ of (\ref{pb:relaxed_outer_approx_k}) are getting unfeasible.
The bi-level procedure is detailed in Algorithm \ref{alg:bilevel_oa} (see Appendix \ref{apdx:bilevel_algorithm}), and illustrated as a workflow in Fig. \ref{fig:workflow}. 

The proposed algorithm leverages the use of post-optimal sensitivities by using them to define supporting hyperplanes of $\Psi$. 
These hyperplanes bound the convex hull of the slave problem.
It is worth to note that the number of constraints involved in the master problem (\ref{pb:binary_bendersDecomposition_up}) reduces to the $k$ linearizations of $\Psi$ from the $(k)$ outer iterations, in addition to the $(k)$ linear equality constraints involved in the definition of $C^{n \times p}$.
Indeed, the OA algorithm is used to solve the master problem (\ref{pb:binary_bendersDecomposition_up}), so that all the structural sizing constraints are handled by the slave problem (\ref{pb:binary_parameterized_continuous_opt}).
Hence, the MILP problem (\ref{pb:relaxed_outer_approx_k}) counts only $k + n$ linear constraints (including equality constraints from $C^{n \times p}$), compared to the $k \times ( n \times m + d + n)$ (constraints $\vec s$, $\vec \delta$ and equality constraints from $C^{n \times p}$).

In industrial cases where the number of structural elements $n$ can reach $5000$ elements (e.g., for a fuselage), and the number of constraints $m$ per structural element is about 10. The problem (\ref{pb:relaxed_outer_approx_k}) can thus involve several millions of constraints.
This could induce high computation time \citep{Benson1991,Stolpe2018} when solving the problem (\ref{pb:relaxed_outer_approx_k}).

Furthermore, two interesting properties about the OA algorithm efficiency have been introduced in \citep{Fletcher1994}.
These properties also apply to the proposed methodology, that falls in the theoretical frame of the OA algorithm. The first property (see \cite[Theorem 2]{Fletcher1994}) states that if  $\Psi$ is convex, then Algorithm \ref{alg:bilevel_oa} converges, in a finite number of steps, at an optimal solution of (\ref{pb:binary_bendersDecomposition_up}). If $\Psi$ is linear, then Algorithm \ref{alg:bilevel_oa} trivially converges to the solution of (\ref{pb:binary_bendersDecomposition_up}) in one iteration.
We note that, although, the convexity assumption cannot be verified for general structural optimization, the proposed algorithm can be used independently of such assumption. In the next section, we will show the performance of the proposed method on practical structural optimization test cases.

\section{Numerical results}\label{sec:numerical_results_oa}

In the present section, the proposed methodology will be applied to three different test cases: (i) the well-known 10-bar truss structure \citep{Haftka} adapted in \citep{Merval2008}, (ii) a 2D scalable cantilever structure \citep{Luis2018}, and (iii) a 120-bar dome truss structure \citep{Saka1992}. 
The third test case aims at demonstrating the efficiency of our methodology on complex structures with large number of categorical choices.

In this paper, we will consider problems with four different structural constraints per element (i.e., $m=4$). In this case, one has two constraints in tension and compression, given by, respectively :
{\small
\begin{eqnarray}\label{allowable_constraints1}
s_{i1}(\vec{a}_i,c,\vec{\Phi}_i(\vec{a},\vec{B})) &:=& \frac{\vec{\Phi}_i(\vec{a},\vec{B})}{\vec{a}_i} - \sigma^{t}(c) \\ \label{allowable_constraints2}
s_{i2}(\vec{a}_i,c,\vec{\Phi}_i(\vec{a},\vec{B})) &:=& - \frac{\vec{\Phi}_i(\vec{a},\vec{B})}{\vec{a}_i} - \sigma^{c}(c)
\end{eqnarray}
}
with $\sigma^{t}(c) \in \mathbb{R}$ the stress limit in tension and $\sigma^{c}(c) \in \mathbb{R}$ the stress limit in compression, for a material choice $c \in \{1, \dots, p\}$.
The two other constraints are the Euler and local buckling constraints, respectively, given by 
{\small
\begin{eqnarray}\label{buckling_constraints1}
s_{i3}(\vec{a}_i,c,\vec{\Phi}_i(\vec{a},\vec{B}))&:=& -\frac{\vec{\Phi}_i(\vec{a},\vec{B})}{\vec{a}_i} - \frac{\pi^2 E(c) I(\vec{a}_i,c)}{\vec{a}_i \vec{L}_i^2} \\ \label{buckling_constraints2}
s_{i4}(\vec{a}_i,c,\vec{\Phi}_i(\vec{a},\vec{B}))&:=& -\frac{\vec{\Phi}_i(\vec{a},\vec{B})}{\vec{a}_i} - \frac{4 \pi^2 E(c)\mathcal{K}^2(c)}{12 (1 - \nu^2(c))}, 
% \left(\frac{x^{(1)_i}{x^{(2)_i} \right)
\end{eqnarray}
}
with $E(c)$, $I(\vec{a}_i, c)$, and $\nu(c)$ are respectively the Young's  modulus, the quadratic moment of inertia and the Poisson's ratio of the material for element $i$, given the choice $c \in \{1, \dots, p\}$. The ratio between cross-section internal sizes, depending on the stiffener profile, is given by $\mathcal{K}(c)$. $\ell_i$ denotes the length of element $i$. The local buckling constraint is introduced to prevent buckling of plate-like elements in the cross-section. It compares the stress value in the considered member with the elastic critical stress value for plate buckling.
The derivatives of the weight function and the constraints (with respect to the areas $\vec a$) are obtained by applying the chain-rule theorem (see Appendix~\ref{apdx:derivatives} for more details).

\subsection{Implementation details}\label{ssec:benchmark_solvers_oa}

Algorithm \ref{alg:bilevel_oa} was implemented using the Generic Engine for MDO Scenarios (GEMSEO) \citep{Gallard2018} in Python.
The continuous non-linear optimization problems (i.e., evaluations of $\Psi$) are solved with the Method of Moving Asymptotes (MMA) \citep{Svanberg2002} as implemented in the nonlinear-optimization (NLOPT) package \citep{Johnson}. The MMA solver is capable of handling non-linear continuous optimization problems with inequality constraints. 
The mixed integer linear optimization problems are solved with a branch and cut implemented as the coin-or branch and cut (coin-or/Cbc) in \citep{Johnforrest2018}.
All the default parameters are kept unchanged except the tolerance on the objective function which is set to $10^{-6}$ $kg$. 
In what comes next, the resulting implementation of Algorithm \ref{alg:bilevel_oa} will be called \textsf{Bi-level OA}.

Four solvers will be compared to \textsf{Bi-level OA}. The first solver is a baseline solver where we proceed with a full enumeration of continuous optimizations w.r.t. $\vec a$; see problem (\ref{pb:binary_parameterized_continuous_opt}). At each iteration, all the available choice in $\mathcal C^{(n,p)}$ are tested. The resulting solution will be denoted as \textsf{Baseline}. 
The second solver in the comparison, is a hybrid branch-and-bound \citep{Barjhoux2018} and will be noted \textsf{h-B\&B}. This solver is based on the usual branch-and-bound algorithm where a specific bound method is adapted to tackle the mixed categorical problem. The procedure involves a continuous relaxation problem formulation to compute lower bounds.
In the case where these problems are convex with respect to the sizing variables, the solvers \textsf{Baseline} and \textsf{h-B\&B} are ensured to return the global optimum of the overall problem. The third solver used in the comparison (will be referred as \textsf{Genetic}) is a genetic algorithm \citep{Deb1998} where we used the implementation given by Distributed Evolutionary Algorithms in Python (DEAP) toolbox \citep{DEAP}. Due to the stochastic nature of \textsf{Genetic}, we run ten times the optimizer and keep only the obtained results of the best run.
The fourth solver will be noted \textsf{Bi-level}, it is the bi-level algorithm proposed by \citep{Barjhoux2020}. The \textsf{Bi-level} solver is based on a similar bi-level paradigm as used in \textsf{Bi-level OA}. The main difference lies in the master problem formulation where, in the \textsf{Bi-level}, we minimize a first order-like approximation. For all the solvers \textsf{Bi-level}, \textsf{h-B\&B} and \textsf{Bi-level OA}, we use the MMA method from NLOPT to solve the slave problem.

The computation effort of a given solver will be measured by counting the number of structural analyses (noted \#FEM) including those required by the computation of the gradients (when needed). The obtained optimal weights (by each solver) will be noted $w^*$, the latter will allow us to evaluate the quality of the optima found by each optimizer.
We note also that in our setting, the \textsf{Baseline} solution can be seen as the best known categorical choices for the corresponding problem instance. We note that the quality of the \textsf{Baseline} solutions (being global optima or just local ones) is depending on the practical capabilities of the NLP solver to find a global optimal solution to the problem \eqref{pb:binary_parameterized_continuous_opt}. For this reason, in cases where \eqref{pb:binary_parameterized_continuous_opt} is not convex, the \textsf{Baseline} solver may not guarantee to provide the global optimum. 
However, in all our numerical tests, we observe that the \textsf{Baseline} results (when available) give the best weight values. For that reason, we decided to evaluate how far the categorical choices  are from the \textsf{Baseline} optimal choices. This information is displayed using the Hamming distance (noted $d_h$) where we will count the number of structural elements that have an optimal choice different to the \textsf{Baseline} categorical choices, i.e., $$d_h := \text{cardinal} \big\{ i \in \left\{ 1, \dots, n \right\} \mid [\vec{c^*}]_i \neq ~[\vec{c}_{\mbox{opt}}]_i \big\},$$ where $\vec{c}_{\mbox{opt}}$  is the the optimal \textsf{Baseline} catalogs and $\vec{c^*}$ is the optimal catalogs found by the other solvers.

\subsection{An illustrative example: a 2-bar truss structure}\label{ssec:2_bars_example_oa}

\begin{figure*}
     \centering
     \captionsetup{justification=centering}
      \begin{subfigure}[b]{0.325\textwidth}
            \centering
            %\captionsetup{font=small}
  \includegraphics[width=.6\textwidth]{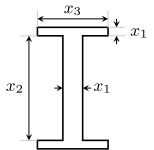}
            \caption{Example of an ``I"-profile  described by 3 geometrical variables.}
        \end{subfigure}
              \begin{subfigure}[b]{0.325\textwidth}
            \centering
            %\captionsetup{font=small}
  \includegraphics[width=.6\textwidth]{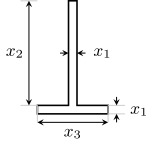}
            \caption{Example of an ``T"-profile, described by 3 geometrical variables.}
        \end{subfigure}
           \begin{subfigure}[b]{0.325\textwidth}
            \centering
            %\captionsetup{font=small}
  \includegraphics[width=.6\textwidth]{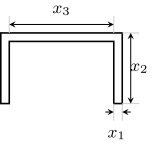}
            \caption{Example of an ``C"-profile, described by 3 geometrical variables.}
        \end{subfigure}
        \caption{Examples of commonly used member profiles in aircraft structural design. The internal geometrical variables are latent variables, scaled by the area of the cross-section.}
        \label{fig:stiffening_principles}
\end{figure*}

\begin{figure}
\centering
%\useexternalfile{1}{2bars_truss}
 \includegraphics[width=.4\textwidth]{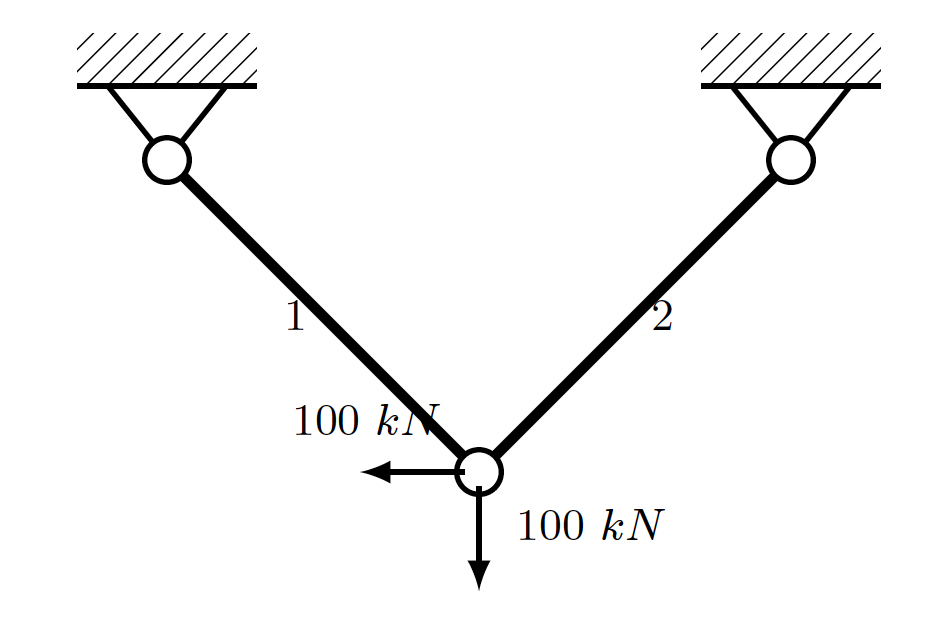}
\caption{A 2-bar truss structure where a downward and leftward load equal to $100~kN$ is applied on the free node.}
\label{fig:2bars_truss_oa}
\end{figure}

To illustrate how the \textsf{Bi-level} method works, we will now describe in details its application to a 2-bar truss structure (see Fig. \ref{fig:2bars_truss_oa}). For this problem, each element can take a value among three possible choices that respectively point to materials AL2139, TA6V and the same ``I''-profile (see Fig. \ref{fig:stiffening_principles}).
The materials properties are listed in Appendix \ref{apdx:120b_case_inputs}.
For this simple case, one has $n=2$, $p=2$, and $\vec B \in C^{2 \times 2}$. 
For all elements, the lower and upper bounds on areas are respectively fixed to $300~mm^2$ and $2000~mm^2$. 
A maximum downward displacement equal to $\bar{\vec{u}} = 7~mm$ is allowed on the only free node of the structure:
\begin{align*}
  \widetilde{\delta} \colon \mathbb{R}^{2} \times \widetilde{\mathcal{C}}^{2,2} &\to \mathbb{R}\\
  (\vec{a}, \vec{B}) &\mapsto \bm P \bm u(\vec{a}, \widetilde{\vec E}(\vec{B})) - \bar{\vec{u}}.
\end{align*}

The \textsf{Bi-level OA} method is initialized with:
\begin{align*}
			\vec{B}^{(0)}=
			\text{vec}\begin{pmatrix} 
			0 & 1 \\
			1 & 0 \\
			\end{pmatrix}
			~~\mbox{and}~~
			\epsilon = 1e{-3}~kg.
\end{align*}
The element 1 is thus made of TA6V, element 2 of AL2139.

\begin{itemize}

	\item \textit{First iteration} $(k=0)$
	
The first iteration $k=0$ starts by solving the primal problem, that reduces to an evaluation of $\widetilde \Psi$ (by solving (\ref{pb:binary_parameterized_continuous_opt})) at the current guess $\vec B^{(0)}$:
\begin{align*}
	U^{(0)} = 5.6~kg,~ \vec a^{(0)} = \left[300.0,~942.8\right]~mm^2
\end{align*}

Then, the gradient $\frac{d\Psi}{d\vec B}\Big|_{\vec B^{(0)}}$ is computed.
To that purpose, the active constraints of the problem (\ref{pb:binary_parameterized_continuous_opt}) at $(\vec a^{(0)}, \vec B^{(0)})$ are the lower bound constraint on the area of structural element 1, and the stress constraint in tension on the second structural element, i.e.,
\begin{align*}
	\ubar{\vec a}_1 - \vec a^{(0)}_1 = 0,~ \vec s_{11}(\vec a^{(0)}, \vec B^{(0)}) = 0.
\end{align*} %\delta(\vec a^{(0)}, \vec B^{(0)}) = 0.
Hence, the sets of active constraints indices are
\begin{align*}
	\mathcal A_{\ubar{\vec a}}^{(0)} = \{1\}, \mathcal A_{\vec s}^{(0)} = \{5\}, ~\mbox{and}~\mathcal A_{\delta}^{(0)} = \mathcal A_{\bar{\vec a}}^{(0)} = \{\emptyset\}.
\end{align*}

%\fi

The gradients of the weight and active constraints w.r.t. $\vec a$ are computed, respectively:
\begin{align*}
\dfrac{\partial \widetilde w}{\partial \vec a}\Bigg|_{\vec{z}^{(0)}} = 
\begin{pmatrix}
			6.26e{-3} \\
			3.96e{-3} \\
\end{pmatrix}
^{\top},
\dfrac{\partial \widetilde{\vec s}_{\mathcal{A}_{\vec s}^{(0)}}}{\partial \vec a}\Bigg|_{\vec{z}^{(0)}} = 
\begin{pmatrix}
			 0.		\\
			-0.16	\\
\end{pmatrix}
^{\top},
\end{align*}
where $\vec{z}^{(0)}=(\vec a^{(0)}, \vec B^{(0)})$ and $\vec I_{\mathcal{A}_{\ubar{\vec a}}^{(0)}} = 
\begin{pmatrix} 
			1 \\
			0 \\
\end{pmatrix}
^{\top}$.
One can see that the gradients of the active constraints are linearly independent. Equation (\ref{eq:lagr_multipliers}) leads to the following linear system (with 2 equations and 2 unknown Lagrange multipliers):
\begin{align*}
     \dfrac{\partial \widetilde{w}}{\partial \vec a}\Bigg|_{\vec{z}^{(0)}} 
          + \left[\vec \lambda^{(0)}_{\mathcal{A}^{(0)}_{\vec s}}\right] \dfrac{\partial \widetilde{\vec s}_{\mathcal{A}^{(0)}_{\vec s}}}{\partial \vec a}\Bigg|_{\vec{z}^{(0)}} 
          + \left[\vec \lambda_{\mathcal{A}^{(0)}_{\ubar a}}^{(0)} \right] \vec I_{\mathcal{A}^{(0)}_{\ubar{\vec a}}}
          = 0.
\end{align*}
Then, as the gradients values are substituted by their value, one deduces  the Lagrange multipliers value:
\begin{align*}
\vec \lambda^{(0)}_{\mathcal{A}^{(0)}_{\vec s}} &= 2.49 ~m.s^2 ~\mbox{and}~
\vec \lambda_{\mathcal{A}^{(0)}_{\ubar a}}^{(0)} &= 6.26e{-3} ~kg/mm^2.
\end{align*}
As a remark, these multipliers illustrate the optimal weight (of the slave problem (\ref{pb:binary_parameterized_continuous_opt})) sensitivity with respect to a perturbation of the constraint on the area lower bound or stress constraint in tension, respectively.

The gradients of the weight and the stress constraints w.r.t. $\vec B$ are computed, respectively:
\begin{align*}
\dfrac{\partial \widetilde w}{\partial \vec B}\Bigg|_{\vec{z}^{(0)}} &= 
\left[1.2, 1.9, 3.7, 5.9 \right], \\
\dfrac{\partial \widetilde{\vec s}_{\mathcal{A}^{(0)}_{\vec s}}}{\partial \vec B}\Bigg|_{\vec{z}^{(0)}} &=
\left[0., 0., 0., -9.5e^{2} \right].
\end{align*}
Thus, using equation \ref{eq:post_opt_sensitivity_oa}, one deduces the gradient of $\Psi$:

{\small
\begin{eqnarray*}
%\begin{split}
 \dfrac{d\Psi}{d\vec B}\Bigg|_{\vec B^{(0)}} &= &
        \dfrac{\partial \widetilde{w}}{\partial \vec B}\Bigg|_{\vec{z}^{(0)}} 
          +  \left[\vec \lambda^{(0)}_{\mathcal{A}^{(0)}_{\vec s}}\right]^{\top} \dfrac{\partial \widetilde{\vec s}_{\mathcal{A}^{(0)}_{\vec s}}}{\partial \vec B}\Bigg|_{\vec{z}^{(0)}} \\
          &=& 	\left[1.2,   1.9,   3.7, -17.7 \right].
\end{eqnarray*}
}
Physically, the values seem to indicate that the optimal weight (of the slave problem (\ref{pb:binary_parameterized_continuous_opt})) is more sensitive to the choices of materials on the second structural element, when compared to the other one.
Indeed, these sensitivities are only valid in a (close enough) neighborhood of $\vec B^{(0)}$.
A change in the active constraint set could occur at intermediate values of $\vec B$.

The history of the previous iterations is updated with $\vec B^{(0)}$ such that:
\begin{align*}
K^{(0)} = \{ \vec B^{(0)} \}.
\end{align*}
The MILP problem (\ref{pb:relaxed_outer_approx_k}) can now be set up.
The solution of this problem provides the new integer candidate solution given by
\begin{align*}
			\vec{B}^{(1)}=
			\text{vec}\begin{pmatrix} 
			1 & 0 \\
			0 & 1 \\
			\end{pmatrix}.
\end{align*}
The optimal objective value is
\begin{align*}
			\eta^{(0)}=-38.71,
\end{align*}
meaning that the difference between the best known guess $U^{(0)}$ and the relaxed problem optimal objective value $\eta^{(0)}$ is lower than the given tolerance $\epsilon$.
Fig. \ref{fig:plot_fig_2b_ite1} shows the supporting hyperplane that provides the feasible set of the MILP problem at the first iteration. The plotted supporting hyperplane, defined over $[0,1]\times[0,1]$, corresponds to the curve surface of the function $$\small{\begin{pmatrix} 
			B_{11}  \\
			B_{21} 
          \end{pmatrix}  \to \Psi(\vec B^{(0)}) + \dfrac{d\Psi}{d\vec B}\Bigg|_{\vec B^{(0)}}^{\top}~\left[\begin{pmatrix} 
			B_{11} \\ 1-B_{11} \\
			B_{21} \\ 1-B_{21} \\
          \end{pmatrix} - \vec B^{(0)}\right].}$$
         We note that one has $U^{(0)}~=
         ~\Psi(\vec B^{(0)})$.

\begin{figure*}
		\captionsetup[subfigure]{justification=centering}
        \centering
        \begin{subfigure}[b]{0.325\textwidth}
            \centering
            %\captionsetup{font=small}
            \includegraphics[width=1\textwidth]{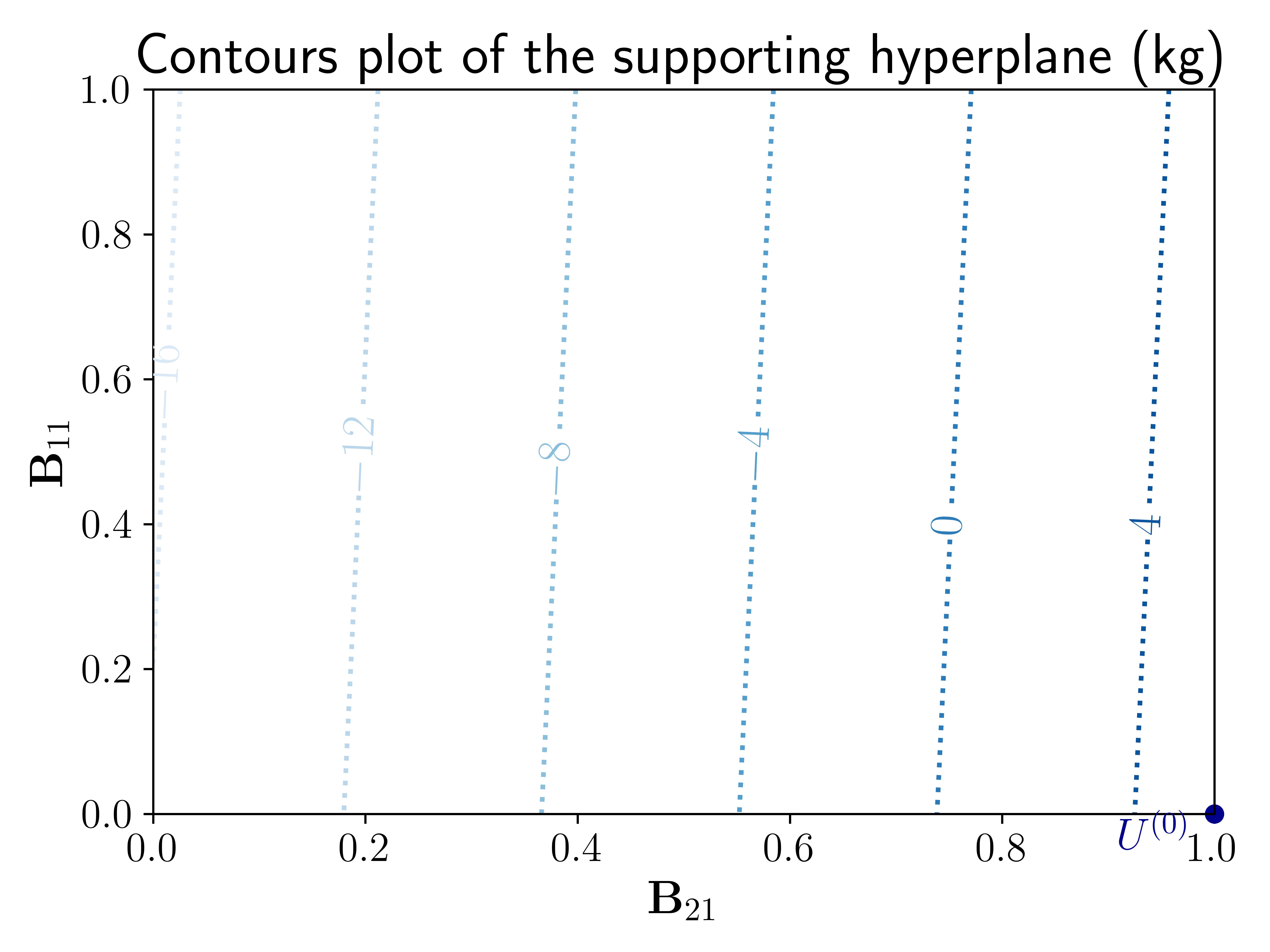}
            \includegraphics[clip, trim=1cm 0cm 2cm 1cm,width=1\textwidth]{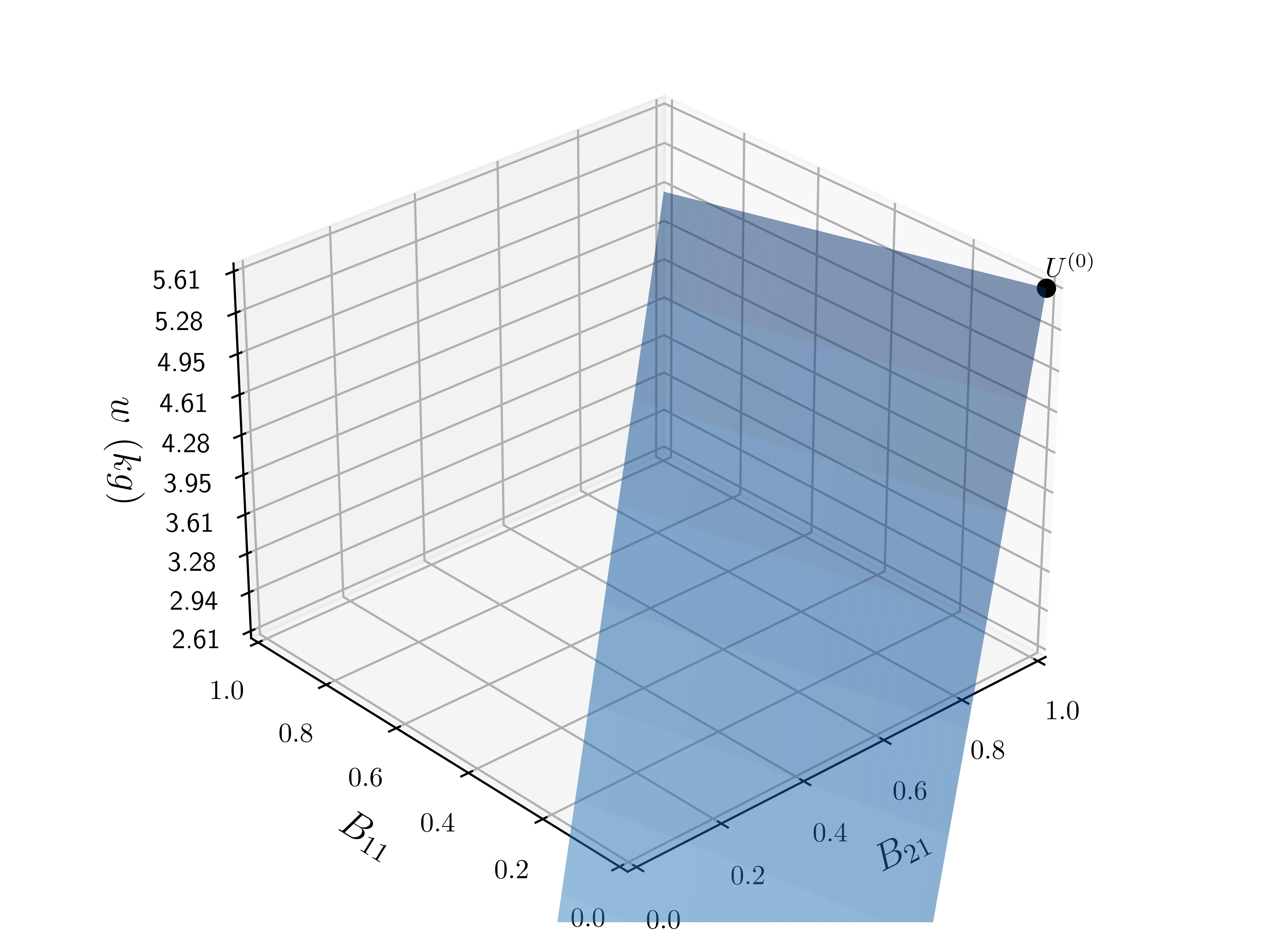}
            \caption{A first supporting hyperplane of $\Psi$ from the first iteration.}
            \label{fig:plot_fig_2b_ite1}   
        \end{subfigure}
        \hfill
        \begin{subfigure}[b]{0.325\textwidth}  
            \centering 
            \captionsetup{font=small}
            \includegraphics[width=1\textwidth]{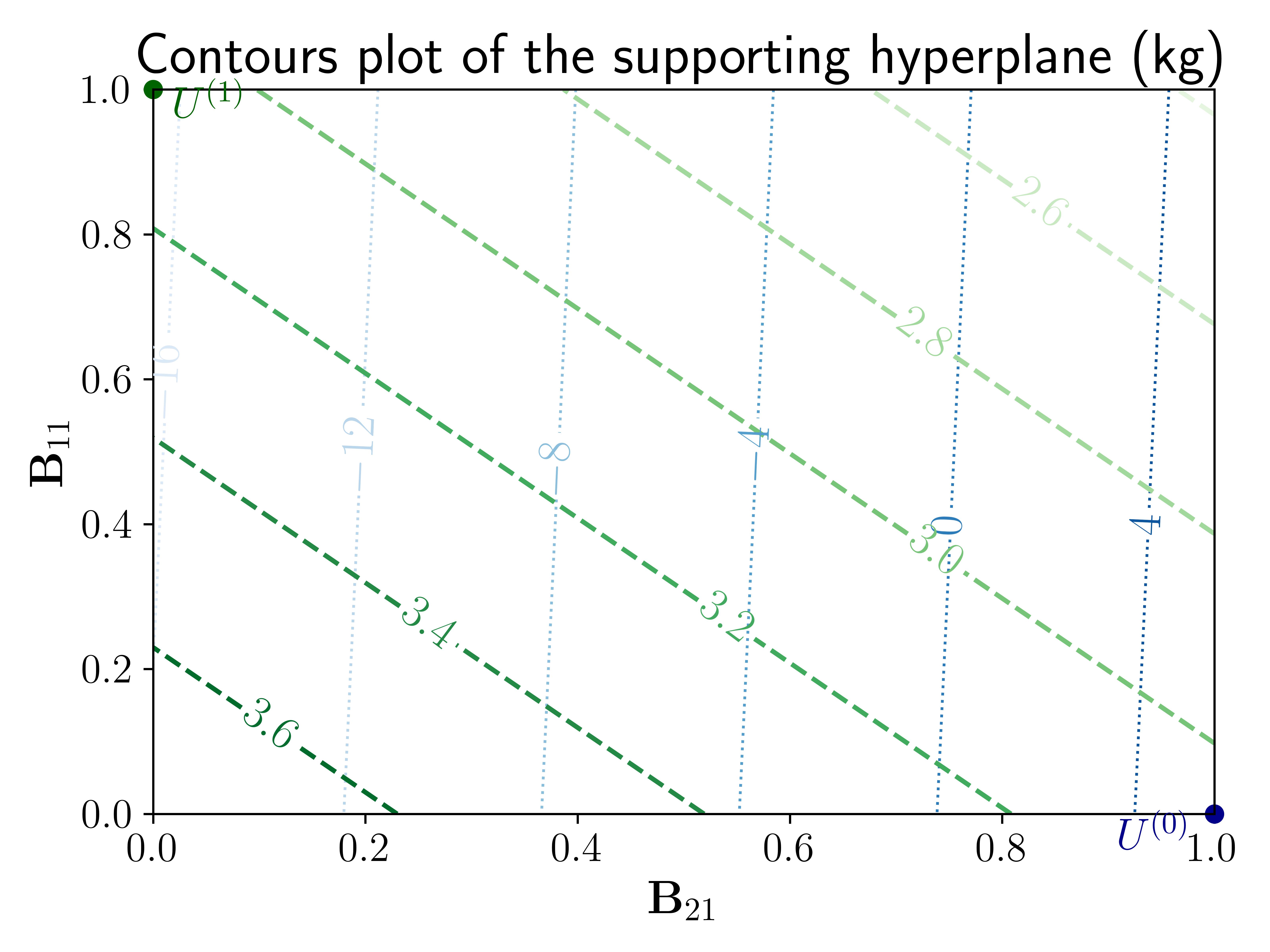}
            \includegraphics[clip, trim=1cm 0cm 2cm 1cm,width=1\textwidth]{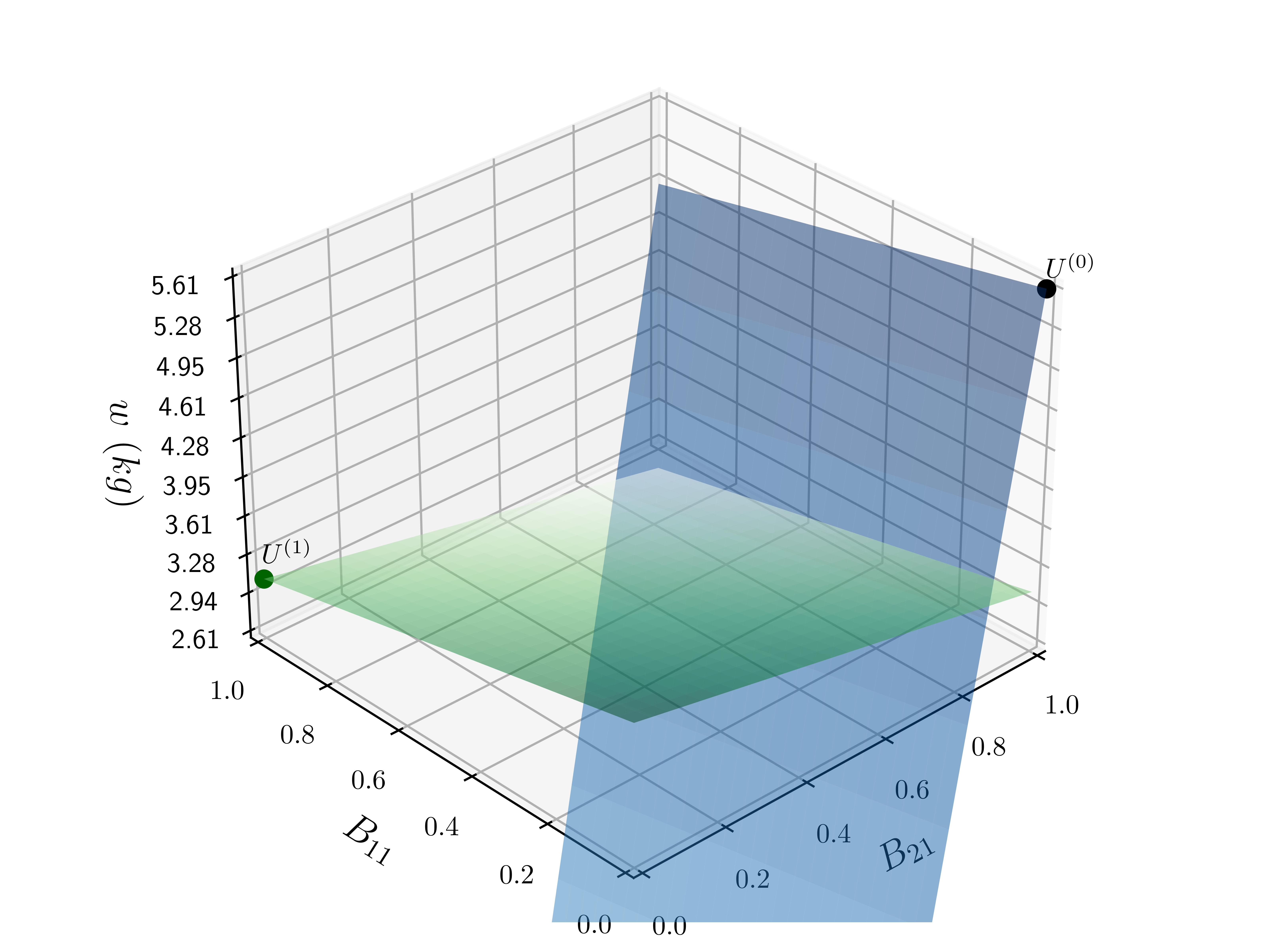}
            \caption{A second supporting hyperplane of $\Psi$ from the second iteration.}
            \label{fig:plot_fig_2b_ite2}
        \end{subfigure}
        \hfill
        \begin{subfigure}[b]{0.325\textwidth}  
            \centering 
            \captionsetup{font=small}
            \includegraphics[width=1\textwidth]{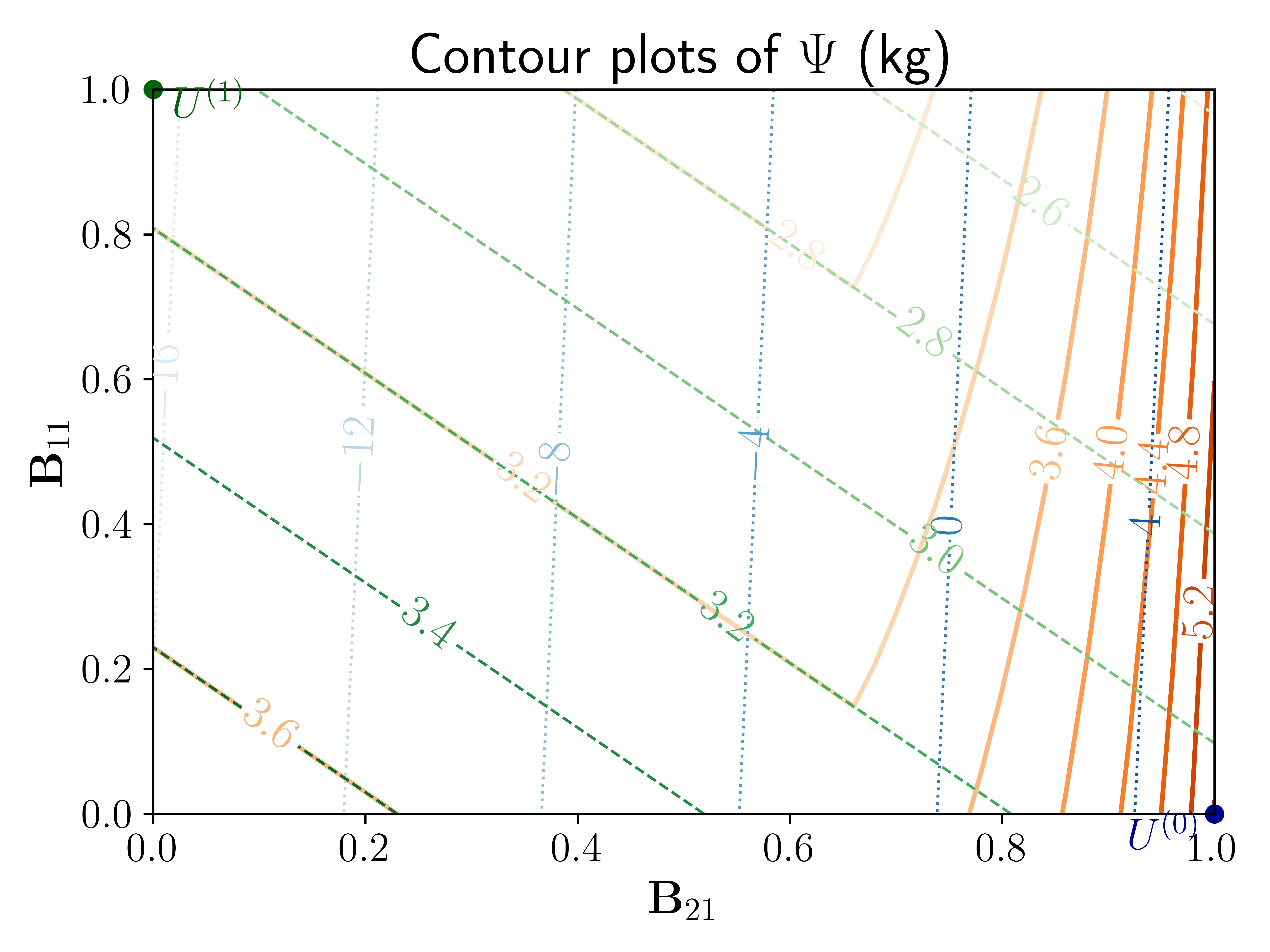}
            \includegraphics[clip, trim=1cm 0cm 2cm 1cm,width=1\textwidth]{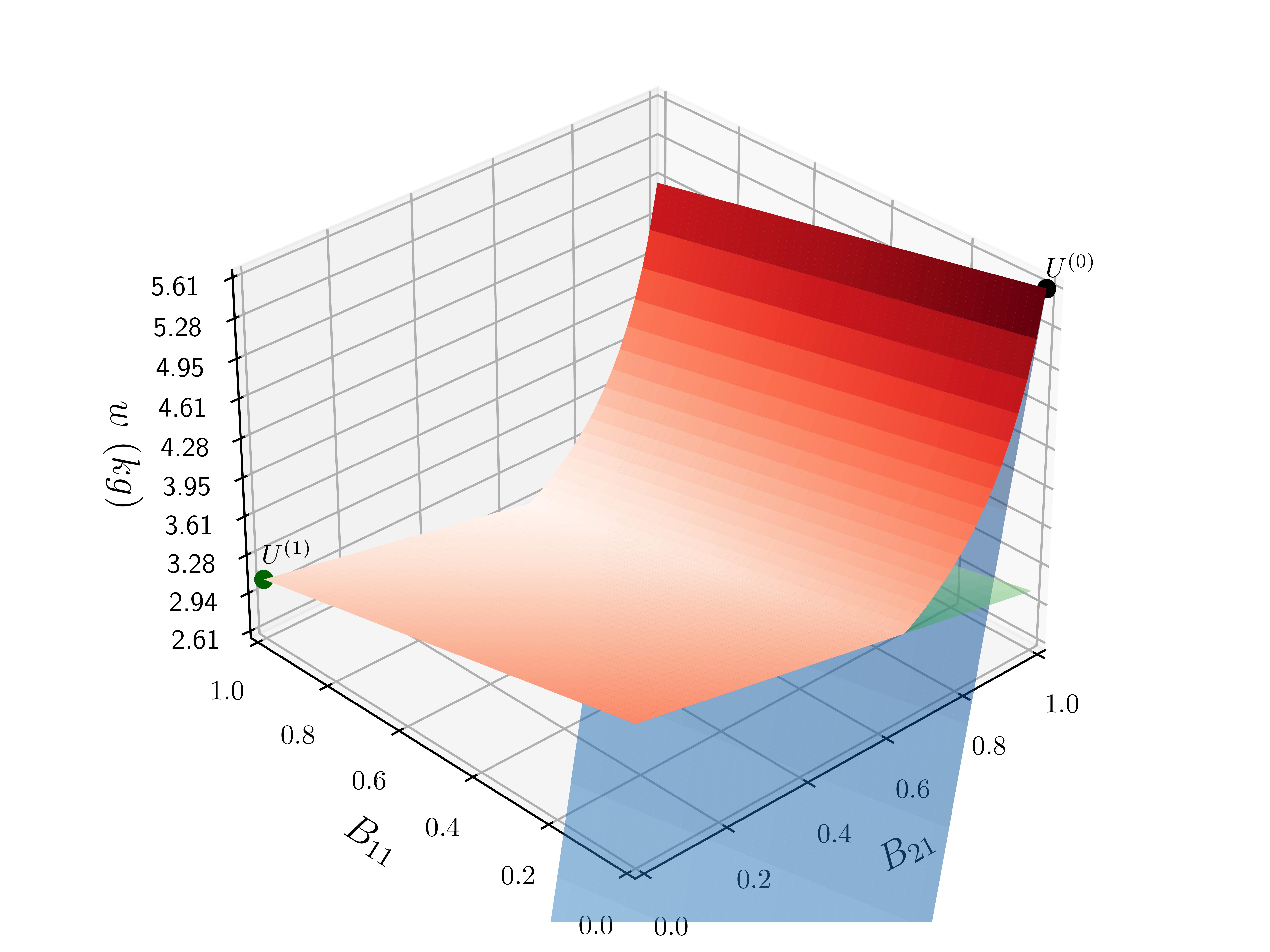}
            \caption{Plot of the surface $\Psi$ for every value of $\vec{B} \in \widetilde C^{2 \times 2}$.}
            \label{fig:plot_fig_2b_finish}
        \end{subfigure}
        \caption{Iterations of a 2-bar truss structure optimization example. The supporting hyperplanes at $\Psi(\vec B^{(0)})$ (resp., $\Psi(\vec B^{(1)})$) are built using the slope $\dfrac{d\Psi}{d\vec B}$ taken at $\vec B^{(0)}$ (resp., $\vec B^{(1)}$).}
        \label{fig:plot_fig_2b}
\end{figure*}

	\item \textit{Second iteration} $(k=1)$

The second iteration starts by solving primal problem, that reduces to an evaluation of $\Psi$ (by solving (\ref{pb:binary_parameterized_continuous_opt})) at the current guess $\vec B^{(1)}$:
\begin{align*}
	U^{(1)} = 3.07~kg,~ \vec a^{(1)} = \left[300.,~300. \right]~mm^2
\end{align*}
Then, similarly to the first iteration of the algorithm, %the gradient $\frac{d\widetilde \Psi}{d\vec B}\Big|_{\vec B^{(1)}}$ is computed.
we estimate the gradient of $\Psi$ with respect to $\vec B$:
\begin{equation*}
\begin{aligned}
        \dfrac{d\Psi}{d\vec B}\Bigg|_{\vec B^{(1)}} 
        &=  \left[1.19, 1.88, 1.19, 1.88 \right].
\end{aligned}
\end{equation*}
The history of the previous iterations is updated with $\vec B^{(1)}$ i.e., $
K^{(1)} = K^{(0)} \cup \{ \vec B^{(1)} \}$ and
the MILP problem (\ref{pb:relaxed_outer_approx_k}) can now be set up, as follows:
\begin{equation}%\tag{MILP($K^{(1)}$)}
\label{pb:relaxed_outer_approx_1}
\begin{aligned}
& \underset{\vec B \in C^{2 \times 2}}{\text{min}} && \eta \\ 
& \underset{}{\text{\textup{subject to}}} && \eta \leq U^{(1)} - \epsilon \\
& && \eta \geq \Psi(\vec B^{(1)}) + \dfrac{d\Psi}{d\vec B}\Bigg|_{\vec B^{(1)}}^{\top}~(\vec B - \vec B^{(1)}) \\
& && \eta \geq \Psi(\vec B^{(0)}) + \dfrac{d\Psi}{d\vec B}\Bigg|_{\vec B^{(0)}}^{\top}~(\vec B - \vec B^{(0)}) 
\end{aligned}
\end{equation}
Fig. \ref{fig:plot_fig_2b_ite2} depicts the result $U^{(1)}$ of the NLP problem (\ref{pb:binary_parameterized_continuous_opt}) solved at $B^{(1)}$, and the associated new hyperplane behaves as an additional constraint for the new MILP problem. 
The optimal objective value is
\begin{align*}
			\eta^{(1)}=3.07,
\end{align*}
that is equal to the best known guess $U^{(1)}$.
This means that the current lower bound of the problem solution is now equal to its current upper bound.
The problem (\ref{pb:relaxed_outer_approx_1}) is thus infeasible, due to the first constraint violation.
The solution found during this second iteration is the optimal solution.

The algorithm then stops, and the solution is such that:
\begin{align*}
\widetilde{w}^* &= U^{(1)} = 3.07~kg	\\ \nonumber
\vec{a}^{*} &= \vec{a}^{(1)} = \left[300, 300 \right]~mm^2, \\ \nonumber
\vec{B}^* &= \vec{B}^{(1)} =
			\text{vec}\begin{pmatrix} 
			1 & 0 \\ \nonumber
			0 & 1
			\end{pmatrix}.
\end{align*}
In other words the optimal material for elements 1 and 2 is AL2139 and TA6V, respectively.
\end{itemize}

Fig. \ref{fig:plot_fig_2b_finish} depicts the landscape of the function $\Psi$ with respect to $\vec{B} \in C^{2 \times 2}$. One can see that 
the admissible solutions of the problem \ref{pb:OverallPbBinary} applied to the 2-bar truss example are the four points at the boundary of $\Psi$ where $B_{11}$ and $B_{21}$ take integer values. 
The optimal solution then corresponds to the point with the lowest value. 
\begin{remark}
At the end of the optimization process, the  optimal values for $\eta$ coincides with the  optimal value of $\Psi$. However, during the minimization process, the obtained values of $\eta$ 
  do not have necessarily a  physical meaning; their values depends on the quality of the approximation provided by the convex hull based on the hyperplanes. 
  For instance, during the value for $\eta$ is negative because the supporting hyperplane is not a good approximation for the function $\Psi$. The convex hull is then refined iteratively (by including new hyperplanes) until $\eta$ corresponds to the value of $\Psi$ at the final solution. 
\end{remark}
\begin{remark} For the \textsf{Bi-level OA} solver, the total computational cost is reduced to the computational effort required to solve 2 NLP problems and 2 MILP problems.
Solving the same illustrative problem by enumeration (\textsf{Baseline}) would have require $2^2$ NLP optimization problems. The \textsf{Bi-level} algorithm as proposed in \cite{Barjhoux2020} would required solving 6 NLP optimization problems.
\end{remark}

\subsection{A 10-bar truss structure}\label{ssec:10bar_truss_oa}

\begin{figure}
%\centering
%\useexternalfile{0.6}{10bars_truss_oa}
 \includegraphics[width=.4\textwidth]{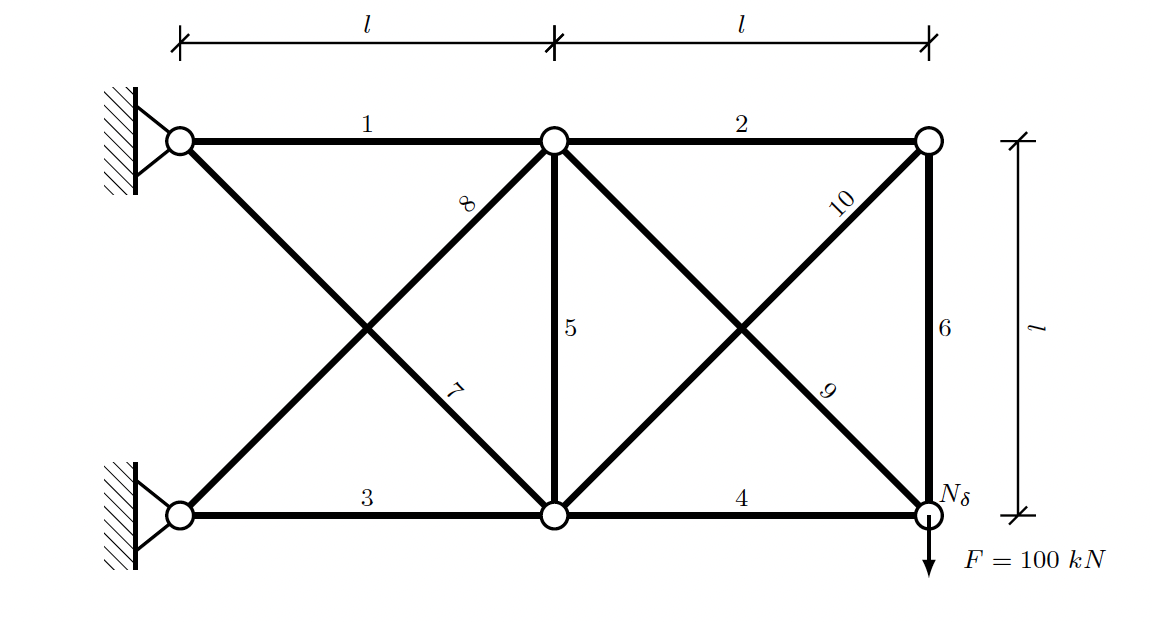}
\caption{10-bar truss, seen as a scalable 2D cantilever problem with 2 blocks.}
\label{fig:10bars_truss_oa}
\end{figure} 

The 10-bar truss problem \citep{Haftka} is used to solve the mixed categorical-continuous optimization problem by enumeration, \textsf{Bi-level} or hybrid branch and bound (\textsf{h-B\&B}) \citep{Barjhoux2017}. 

The 10-bar truss problem is illustrated Fig. \ref{fig:10bars_truss_oa}. A downward load $F = 100~kN$ is applied vertically on node $N_{\delta}$. A constraint on displacements is applied on the same node. Five cases with different bounds values  $\bar{\vec{u}}$ on displacements are considered. For each of these cases, the displacements constraint is applied on node $N_{\delta}$. 
Each structural element is also subjected to the stress constraints given by \eqref{allowable_constraints1}, \eqref{allowable_constraints2}, \eqref{buckling_constraints1} and \eqref{buckling_constraints2}.
The lower bounds, the upper bounds, and the initial areas are fixed to $100  ~mm^2$, $1300 ~mm^2$ and $1300 ~mm^2$, respectively.
Catalogs 1 and 2 point to materials AL2139 and TA6V, respectively. Materials properties are listed in Appendix \ref{apdx:120b_case_inputs}.
For this simple case, one has $n=10$, $p=2$, and $\vec B \in \mathcal C^{10,2}$. 

Table \ref{tab:10bar_truss_OA} depicts the obtained results on a 10-bar truss mixed optimization using 5 different values of constraint on displacements. 
In all these cases, as shown by the Hamming distance $d_h$ and the optimal weights $w^*$, the solutions found by \texttt{Baseline}, \textsf{h-B\&B}, \textsf{Bi-level} and \textsf{Bi-level OA} solvers are identical.
The  optimal solutions returned by the \textsf{Genetic} solver are not as good as the optimal weights found by the the rest of the solvers. In fact, although sometimes the \textsf{Genetic} solver is able to find the optimal catalogs (since $d_h=0$), the continuous variables are not well handled (since $w^*$ is higher compared to the other solvers).
The displacement constraint are active in all the cases; as far as the displacement constraint becomes more stringent, the material choice goes to the stiffest one despite of its high density. In our experiments, the optimal solutions of cases with maximum displacements equal to $18 mm$ and $17 mm$ contain indeed only TA6V material.
Regarding the other constraints, the Euler buckling constraints were active on elements 10 and 8 for both test cases cases with a maximum displacement of $20mm$ and $22mm$.
The constraints were also active in all cases, but on different elements depending on the bound value on displacements. Namely, for the cases with a maximum displacement equals to $19mm$, $20mm$ and $22mm$, the stress constraints were active for the elements 2, 6 and 9. For the case $18mm$, the same constraint was active for the elements 1, 2, 6, 7 and 9; for the case $17mm$ the constraint was active for elements 1 and 7. Unlike the displacement constraints,  in the provided examples, the local buckling constraints were not active at the solution. It seems that in our setting such constraints do not have a significant role in the optimization process.

\begin{table*}
\centering
\ra{1.5}
%\begin{adjustbox}{tabular=lll,center}
\centerline{
\begin{tabular}{@{\extracolsep{2pt}}cccccccccccc@{}}
\toprule
\multirow{2}{*}{$\bar{\vec{u}}~(mm)$} & \multicolumn{2}{c}{\textsf{Baseline}} & \multicolumn{2}{c}{\textsf{h-B\&B}} & \multicolumn{2}{c}{\textsf{Genetic}} & \multicolumn{2}{c}{\textsf{Bi-level}} & \multicolumn{2}{c}{\textsf{Bi-level OA}}\\ 
\cmidrule{2-3} \cmidrule{4-5} \cmidrule{6-7} \cmidrule{8-9} \cmidrule{10-11}
 & $\vec c^* = \vec B \vec \gamma $         & $w^* (kg)$   & $d_h$       & $w^* (kg)$   & $d_h$   & $w^* (kg)$   & $d_h$ & $w^* (kg)$ & $d_h$ & $w^* (kg)$ \\
                       \midrule
-22    & [2,2,1,1,1,2,2,1,2,1]           & 12.988   & 0        & 12.988     &  0   & 13.283  &    0         & 12.988   &   0   &  12.988   \\
-20    & [2,1,1,1,1,1,2,1,1,1]           & 13.996   & 0        & 13.996     &  0   & 14.423  &    0         & 13.996   &   0   &  13.996   \\
-19    & [2,1,1,1,1,1,2,1,1,1]           & 14.570   & 0        & 14.570     &  0   & 14.802  &    0         & 14.570   &   0   &  14.570   \\
-18    & [1,1,1,1,1,1,1,1,1,1]           & 15.175   & 0        & 15.175     &  2   & 15.642  &    0         & 15.174   &   0   &  15.174   \\
-17    & [1,1,1,1,1,1,1,1,1,1]           & 15.912   & 0        & 15.912     &  3   & 16.258  &    0         & 15.912   &   0   &  15.912   \\
\bottomrule
\end{tabular}
}
\caption{\label{tab:10bar_truss_OA}Results of the 10-bar truss testcase with 5 different values of constraint on displacements. Comparison between the \textsf{Bi-level OA}, \textsf{Bi-level}, the \textsf{Baseline} solutions obtained by enumeration of the $2^{10}$ continuous optimizations, \textsf{h-B\&B}, and the \textsf{Genetic} algorithm. The catalog 1 corresponds to material AL2139 and catalog 2 to TA6V.}
%\end{adjustbox}
\end{table*}

\subsection{Scalability of our algorithm}

\subsubsection{Scalability with respect to the number of elements}

\begin{figure}
%\centering
%\useexternalfile{0.6}{xbars_truss}
 \includegraphics[width=.5\textwidth]{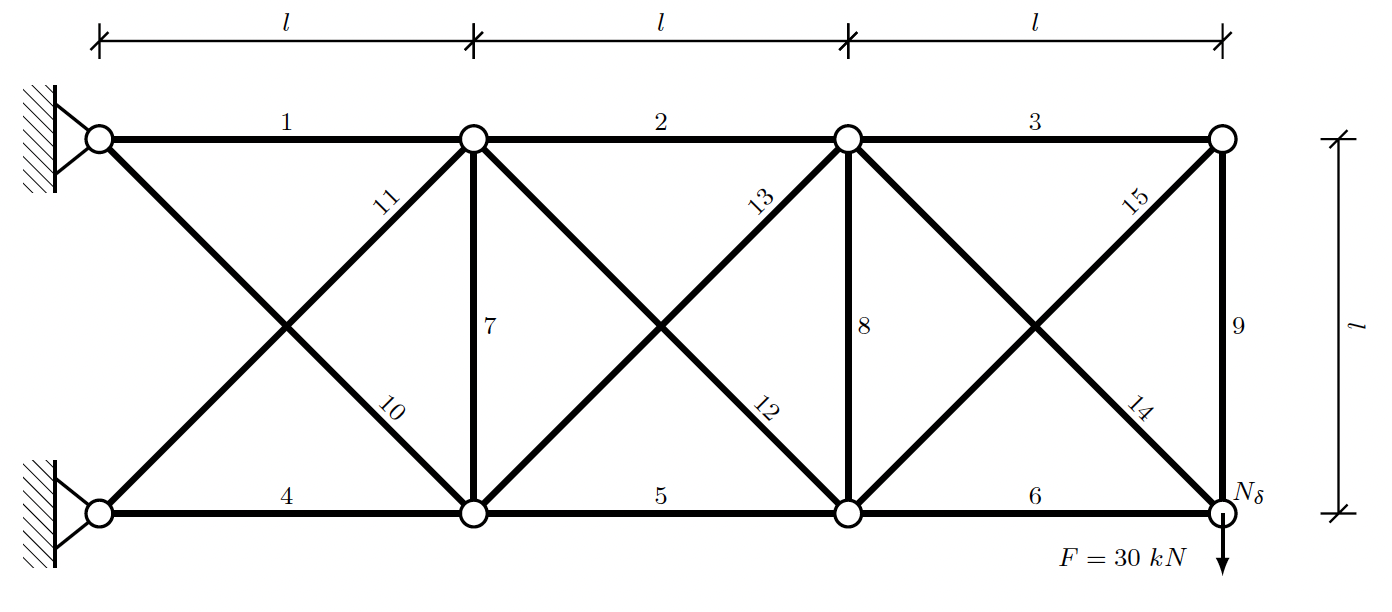}
\caption{An example of 2D cantilever problem with 3 blocks.}
\label{fig:xbars_truss_oa}
\end{figure} 

The objective of this test case is to describe the evolution of the computational cost with respect to the number of structural elements. It has been used in the literature to demonstrate the scalability of algorithms, see for instance \citep{Luis2018}. Each block is composed of 4 nodes that are linked by 5 bars. An example of a parametric 2D cantilever structure with 3 blocks is given in Fig. \ref{fig:xbars_truss_oa}. Table \ref{tab:scaling_results_elts_oa} presents the results obtained with structures composed of 1 to 10 blocks. In all cases, a force load $F = 30~kN$ is applied on the node $N_{\delta}$.
The lower bounds, the upper bounds, and the initial areas are fixed to $100  ~mm^2$, $2000 ~mm^2$ and $2000 ~mm^2$, respectively. No constraint on displacements is considered and each structural element is subjected to the stress constraints given by \eqref{allowable_constraints1}, \eqref{allowable_constraints2}, \eqref{buckling_constraints1} and~\eqref{buckling_constraints2}.

\begin{table*}
\centering
\ra{1.4}
\setlength{\tabcolsep}{2pt}
\centerline{
\begin{tabular}{*{22}{c}}\toprule
\multirow{2}{*}{\#bars} & \multicolumn{1}{c}{\textsf{Baseline}} && \multicolumn{4}{c}{\textsf{h-B\&B}} && \multicolumn{4}{c}{\textsf{Genetic}} && \multicolumn{4}{c}{\textsf{Bi-level}} && \multicolumn{4}{c}{\textsf{Bi-level OA}}\\ 
\cmidrule{2-2} \cmidrule{4-7} \cmidrule{9-12} \cmidrule{14-17} \cmidrule{19-22}
                      & $w^* (kg)$  && $d_h$       & $w^* (kg)$  & \#iter  & \#FEM     && $d_h$    & $w^* (kg)$     & \#iter & \#FEM     && $d_h$    & $w^* (kg)$  & \#iter   & \#FEM && $d_h$    & $w^* (kg)$  & \#iter   & \#FEM\\
                      \midrule
5  & 2.56  && 0 & 2.56             & 10                & 1004               && 0 & 2.57  & 32  & 32300  && 0 & 2.56   & 2  & 400   && 0 & 2.56   & 2  & 96   \\
10 & 6.06  && 0 & 6.06             & 26                & 3097               && 1 & 6.14  & 54  & 54500  && 0 & 6.06   & 2  & 792   && 0 & 6.06   & 2  & 181   \\
15 & 10.23 && 0 & 10.23            & 95                & 10907              && 2 & 10.27 & 65  & 65200  && 0 & 10.23  & 4  & 1955  && 0 & 10.23  & 6  & 967  \\
20 & *     && * & 15.33            & 135               & 10315              && * & 15.59 & 73  & 73100  && * & 15.33  & 2  & 1659  && * & 15.33  & 7  & 1023  \\
25 & *     && * & 21.36    	       & 1199              & 610347             && * & 22.06 & 98  & 97700  && * & 21.36  & 3  & 3142  && * & 21.36  & 13 & 2312  \\
30 & *     && * & 28,30            & 4432              & 723388             && * & 28.84 & 129 & 128800 && * & 28.30  & 8  & 10522 && * & 28.30  & 13 & 2991 \\
35 & *     && * & $36,17^{(\ast)}$ & $5793^{(\ast)}$   & $1096968^{(\ast)}$ && * & 37.00 & 189 & 189400 && * & 36.19  & 3  & 5830  && * & 36.19  & 6  & 1496  \\
40 & *     && * & $44,97^{(\ast)}$ & $5570^{(\ast)}$   & $939726 ^{(\ast)}$ && * & 45.64 & 270 & 269800 && * & 44.97  & 7  & 13577 && * & 44.96  & 40 & 11578 \\
45 & *     && * & $54,70^{(\ast)}$ & $4181^{(\ast)}$   & $818455 ^{(\ast)}$ && * & 55.98 & 347 & 346800 && * & 54.71  & 4  & 8531  && * & 54.67  & 20 & 6789  \\
50 & *     && * & $65,35^{(\ast)}$ & $4316^{(\ast)}$   & $717627 ^{(\ast)}$ && * & 67.48 & 561 & 561200 && * & 65.34  & 6  & 14487 && * & 65.34  & 42 & 13290 \\
\bottomrule
\end{tabular}}
\caption{A comparison of the obtained solutions for 10 instances of the scalable 2D cantilever problem are compared, with a varying number of bars (from 5 to 50 bars). We note that when optimizations last more than 24 hours, the solver (\textsf{Baseline}, \textsf{h-B\&B}) is stopped and the current solution (if exists) is marked by $(*)$.}
\label{tab:scaling_results_elts_oa}
%\end{sidewaystable*}
\end{table*}

\begin{figure}
		\captionsetup[subfigure]{justification=centering}
        \centering
        \begin{subfigure}[b]{0.475\textwidth}  
            \centering 
            \captionsetup{font=small}
            \includegraphics[width=1\textwidth]{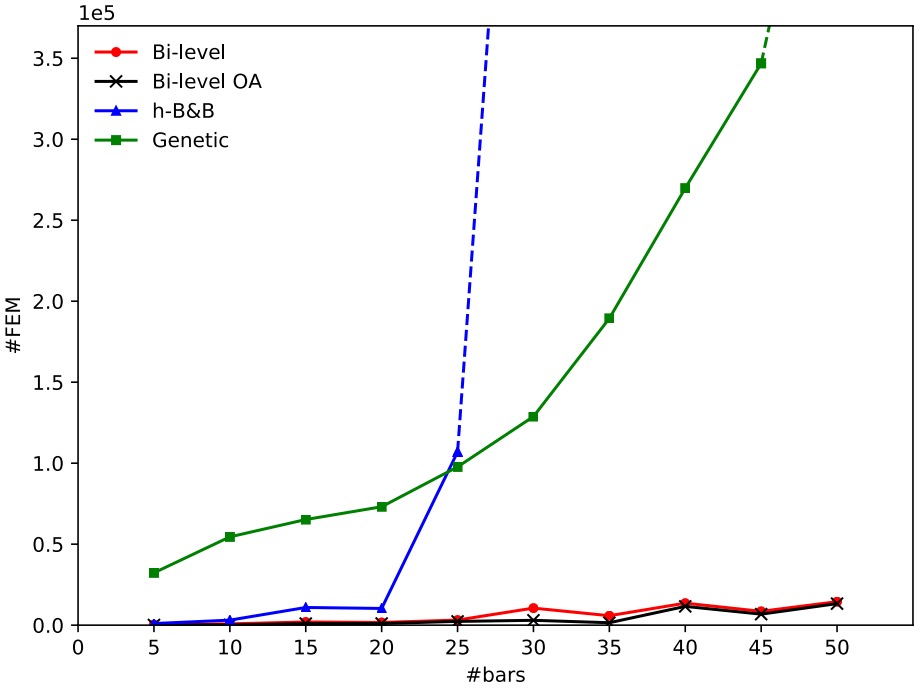}
            \caption{ }
            \label{fig:scaling_elts_oa_full}
        \end{subfigure}
        \hfill
        \begin{subfigure}[b]{0.475\textwidth}
            \centering
            \captionsetup{font=small}
            \includegraphics[width=1\textwidth]{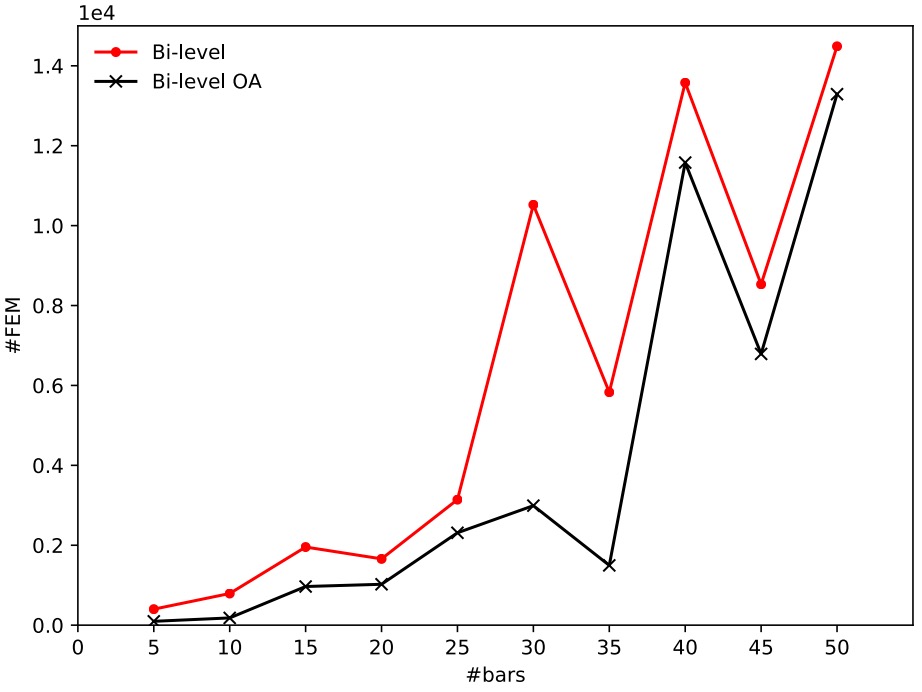}
            \caption{ }
            \label{fig:scaling_elts_oa_bilevels_only}   
        \end{subfigure}
        \caption{Scalability of the \textsf{Bi-level OA} w.r.t. the number of structural elements. The computational cost’s scaling of \textsf{Bi-level OA} and \textsf{Bi-level} with respect to the number of bars is almost linear, compared to the exponential computational cost of the \textsf{h-B\&B} and \textsf{Genetic} solvers shown Fig. \ref{fig:scaling_elts_oa_full}. The high computational cost of the \textsf{h-B\&B} prevents from obtaining a solution for cases greater than 25 elements. The plot on Fig. \ref{fig:scaling_elts_oa_bilevels_only} focuses on a comparison between the computational cost of \textsf{Bi-level} and \textsf{Bi-level OA} only. The computation cost is always lower than the \textsf{Bi-level}. }
        \label{fig:scaling_elts_oa}
\end{figure}

For each of the 10 cases, the results obtained by the \textsf{Bi-level OA} are compared to those obtained with reference solutions (\textsf{Baseline}, \textsf{h-B\&B}) and \textsf{Bi-level} when available. 
First, for the three cases with 5 to 15 elements where a reference solution is available, it can be observed the global solution is found by the \textsf{Bi-level OA}. In these three cases for all the tested solvers the optimal categorical variable values are identical, excepted for the \textsf{Genetic} solver (based on the $d_h$ values).
For cases with more than 15 elements, the optima found by the \textsf{Bi-level OA} are slightly better than those obtained by the \textsf{Genetic} algorithm. The \textsf{h-B\&B} solutions are noted with (*) since they are intermediate solutions: the solver was stopped after 24 hours. The \textsf{Bilevel OA} solutions are very close (difference of $10^{-2}$ kg) to those obtained by the \textsf{h-B\&B}. 
For cases with 40 and 45 elements, the \textsf{Bilevel OA} solutions are slightly lighter than the \textsf{Bilevel}.
Furthermore, the number of analyses required by \textsf{Bi-level OA} is always lower than the number needed by the compared approaches, including \textsf{Bi-level}.
The trends in terms of computational cost with respect to the number of elements are graphically represented in Fig. \ref{fig:scaling_elts_oa}. The cost of the \textsf{Genetic} algorithm dominates the cost of \textsf{h-B\&B} \textsf{Bi-level} and \textsf{Bi-level OA}. As with the \textsf{Bilevel}, the scaling of the \textsf{Bi-level OA} approach is nearly linear when compared to the \textsf{h-B\&B} and \textsf{Genetic} approach. The trends in terms of \textsf{Bi-level OA} computational cost with respect to the number of elements are similar to the \textsf{Bi-level} computation cost.
The observed efficiency makes the proposed approach relevant for higher dimensional problems.

\subsubsection{Scalability with respect to the number of catalogs}

The objective of this test case is to describe the computational cost scaling with respect to the number of categorical choices.
The test case is the same 10-bar truss case presented in Section \ref{ssec:10bar_truss_oa}, with a constraint on displacements such that $\bar{\vec{u}} = 10~mm$.
For this simple case, one has the number of structural elements fixed to $n=10$, but $p$ is varying from 5 to 90 catalogs.
Each catalog is defined as a combination of different materials among AL2139, AL2024 and TA6V, with the member profiles I, T and C. For each member profile we consider using $10$ different sizes.
The material properties, the catalogs and the member profiles are listed in Appendix \ref{apdx:120b_case_inputs} (see, Tables \ref{tab:materialDefinition}, \ref{tab:catalogs} and \ref{tab:profilesDefinition}).
Thus, by scaling the number of catalog choices, the full number of available categorical choices will range from $10^4$ to $10^{90}$.

\begin{figure}
\captionsetup{width=0.5\textwidth}
%\hspace*{-1.5cm}
\begin{center}
\includegraphics[clip, trim=0cm 0cm 0cm
   0cm, scale=0.26]{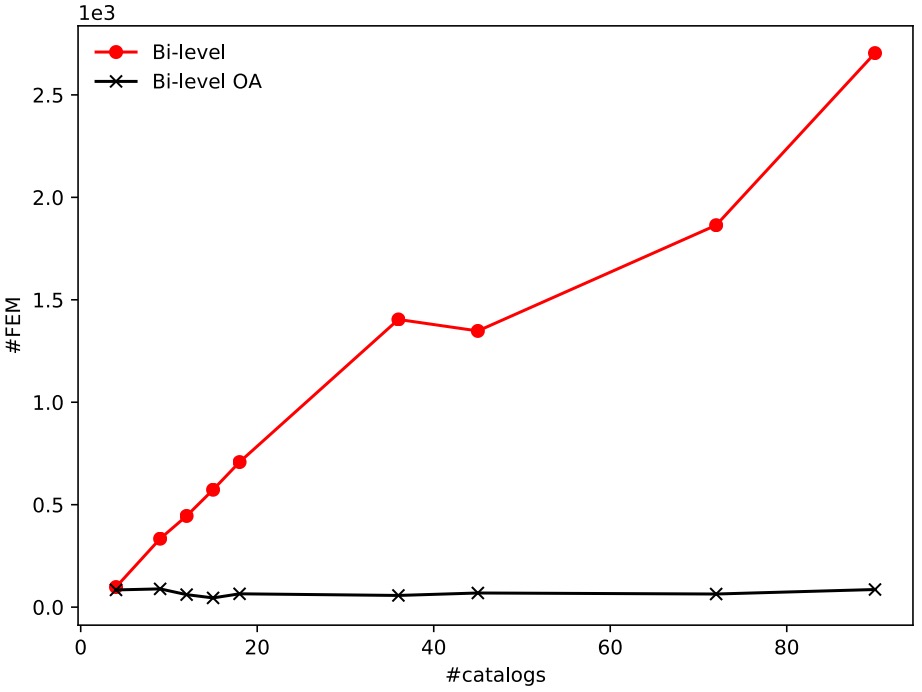}
\end{center}
\caption{Scalability of the \textsf{Bi-level OA} w.r.t. the number of catalogs. The computational cost’s scaling of \textsf{Bi-level OA} with respect to the number of catalogs is nearly independent from the number of catalogs, compared to the quasi-linear computational cost of the \textsf{Bi-level}.}
\label{fig:scaling_cat_oa}
\end{figure}

Table \ref{tab:scaling_results_cat_oa} presents the results obtained by \textsf{Bi-level OA} and \textsf{Bi-level}.
The \textsf{h-B\&B} failed to solve the problem instances in 24 hours.
The optimal weight, the number of iterations (\#ite), non-linear problems (\#NLP) solved, and the number of individual calls to the structural solver (\#FEM) are compared. First, in terms of \#FEM and \#NLP, the computational cost of \textsf{Bi-level OA} reveals to be almost linear with respect to the number of categorical values when compared to \textsf{Bi-level}.
Furthermore, it is shown that for each case, the optimal weights obtained by both solvers are close (the gap is less than $10^{-3} kg$). This shows that the \textsf{Bi-level} is able to return good quality solutions even in cases with a large scale categorical design space. 
Independently from the solver, it can be remarked that the optimal weights are identical from cases 4 to 36, and 45 to 90.
This is due to the fact that the categorical values introduced in the design space from case 4 to 36 (or from 45 to 90) do not lead to any significant improvement for the optimal weight; the improvement is only observed from case 45 to 72.
The computational cost with respect to the number of catalogs for all the tested solvers is depicted in Fig. \ref{fig:scaling_cat_oa}.

%\begin{landscape}
%\thispagestyle{empty}
%\begin{sidewaystable*}
\begin{table*}
\centering
\ra{1.4}
\setlength{\tabcolsep}{6pt}
\centerline{
\begin{tabular}{*{11}{c}}\toprule
\multirow{2}{*}{\#catalogs} && \multicolumn{4}{c}{\textsf{Bi-level}} && \multicolumn{4}{c}{\textsf{Bi-level OA}} \\ 
\cmidrule{3-6} \cmidrule{8-11}
     && $w^* (kg)$       & \#iter  & \#NLP  & \#FEM     &&  	$w^* (kg)$       & \#iter  & \#NLP  & \#FEM \\
                      \midrule
4  && 12.99            & 3    & 98   & 29245     && 12.99            & 84 & 84 & 8400  \\
9  && 12.39            & 4    & 334  & 98871     && 12.39            & 89 & 89 & 3772  \\
12 && 12.39            & 4    & 445  & 131358    && 12.39            & 61 & 61 & 2583  \\
15 && 12.39            & 4    & 573  & 170407    && 12.39            & 45 & 45 & 1955  \\
18 && 12.39    	       & 4    & 708  & 209201    && 12.39            & 65 & 65 & 2877  \\
36 && 12.39            & 4    & 1404 & 416662    && 12.39            & 57 & 57 & 2232  \\
45 && 12.15	           & 3    & 1348 & 399172	 && 12.15            & 69 & 69 & 2489  \\
72 && 12.15            & 3    & 1864 & 551042    && 12.15            & 64 & 64 & 2898  \\
90 && 12.15            & 3    & 2704 & 799166    && 12.15            & 86 & 86 & 3952  \\
\bottomrule
\end{tabular}}
\caption{A comparison of the obtained solutions for 9 instances of the 10-bar truss problem are compared, with a varying number of catalogs (from 4 to 90 catalogs).}
\label{tab:scaling_results_cat_oa}
%\end{sidewaystable*}
\end{table*}
%\end{landscape}

\subsection{120-bar truss}\label{ssec:120_bars_truss_oa}

\begin{figure}
\centering
%\useexternalfile{0.7}{120bars_truss}
 \includegraphics[width=.5\textwidth]{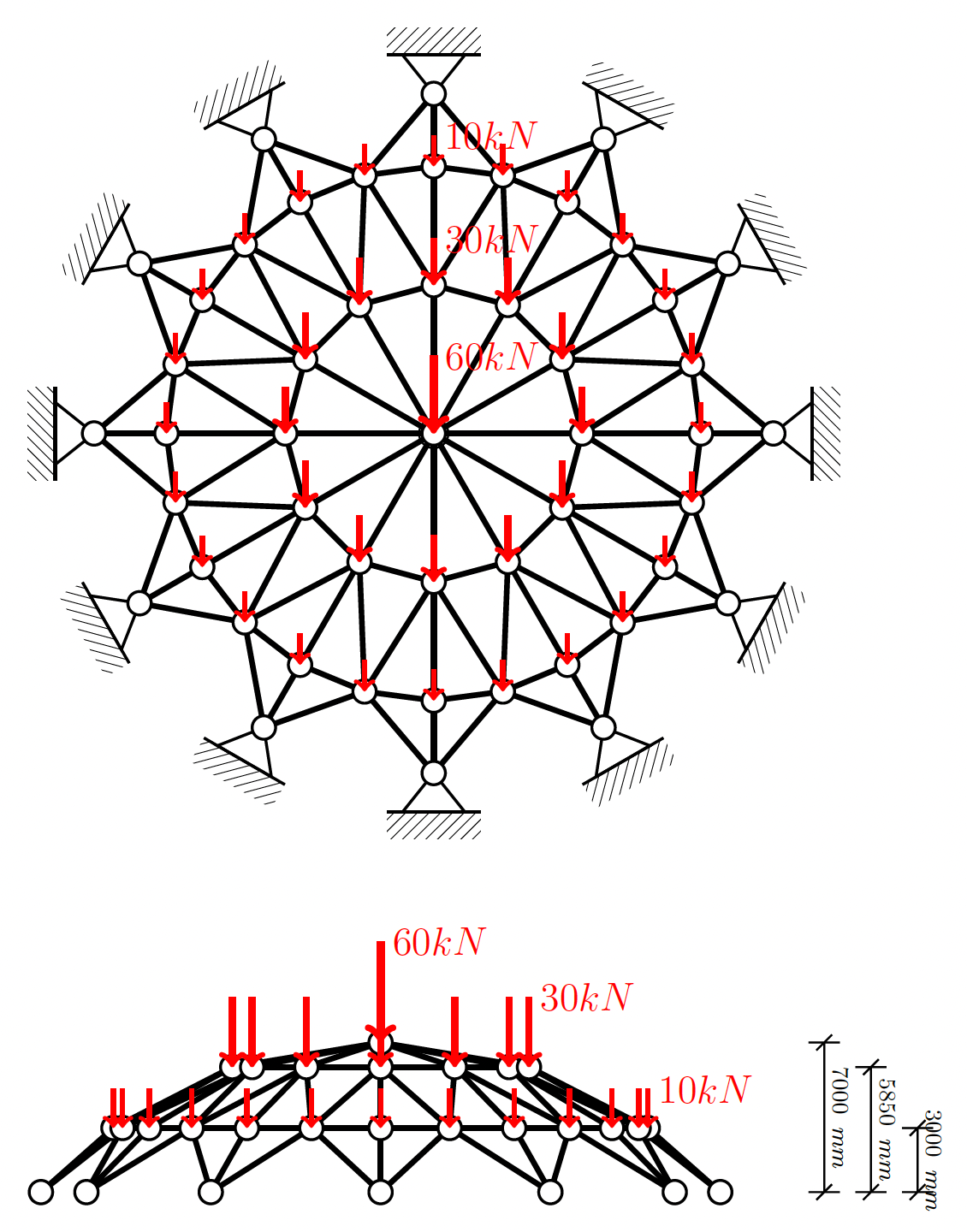}

\caption{Top and side view a 120-bar truss structure. Downward loads with three different magnitudes are applied.}
\label{fig:120bars_truss_oa}
\end{figure} 

In this example, the structure of a 120-bar dome truss \citep{Saka1992} detailed in Fig. \ref{fig:120bars_truss_oa} is considered.
For each element, the categorical variable can take a value among $90$ catalogs. With $n=120$ and $p=90$, the binary design space is $\mathcal{C}^{120\times 90}$. The number of available categorical choices is thus equal to $90^{120}$.
Each catalog is defined as a combination of materials among AL2139, AL2024 and TA6V, with profiles I, T and C (with 10 different sizes for each profile).
The material properties are listed in Table \ref{tab:materialDefinition} in Appendix \ref{apdx:120b_case_inputs}.
The catalogs are listed in Table \ref{tab:catalogs} and the profiles in Table \ref{tab:profilesDefinition}, in Appendix \ref{apdx:120b_case_inputs}. The structure is subjected to a constraint on displacements: a maximum downward displacement of $10~mm$ is allowed on node 1 (i.e., the top of the dome). A downward load of $60~kN$ is applied on this same node, while 12 downward loads of $30~kN$ are applied on nodes 2 to 13 (i.e., inner ring) and $10~kN$ on nodes 14 to 37 (i.e., outer ring). 
For this test case, the lower bounds, the upper bounds, and the initial areas are fixed to $100  ~mm^2$, $6000 ~mm^2$ and $6000 ~mm^2$, respectively. As for the previous tested problems, the areas of all the structural elements are handled as continuous design variables and each element is subjected to the stress constraints  \eqref{allowable_constraints1}, \eqref{allowable_constraints2}, \eqref{buckling_constraints1} and \eqref{buckling_constraints2}. The stress constraints will assess the structural integrity of the truss and avoid buckling in the members.

\begin{figure}
\begin{center}
\includegraphics[clip, trim=1.7cm 0cm 0cm 0cm, scale=0.6]{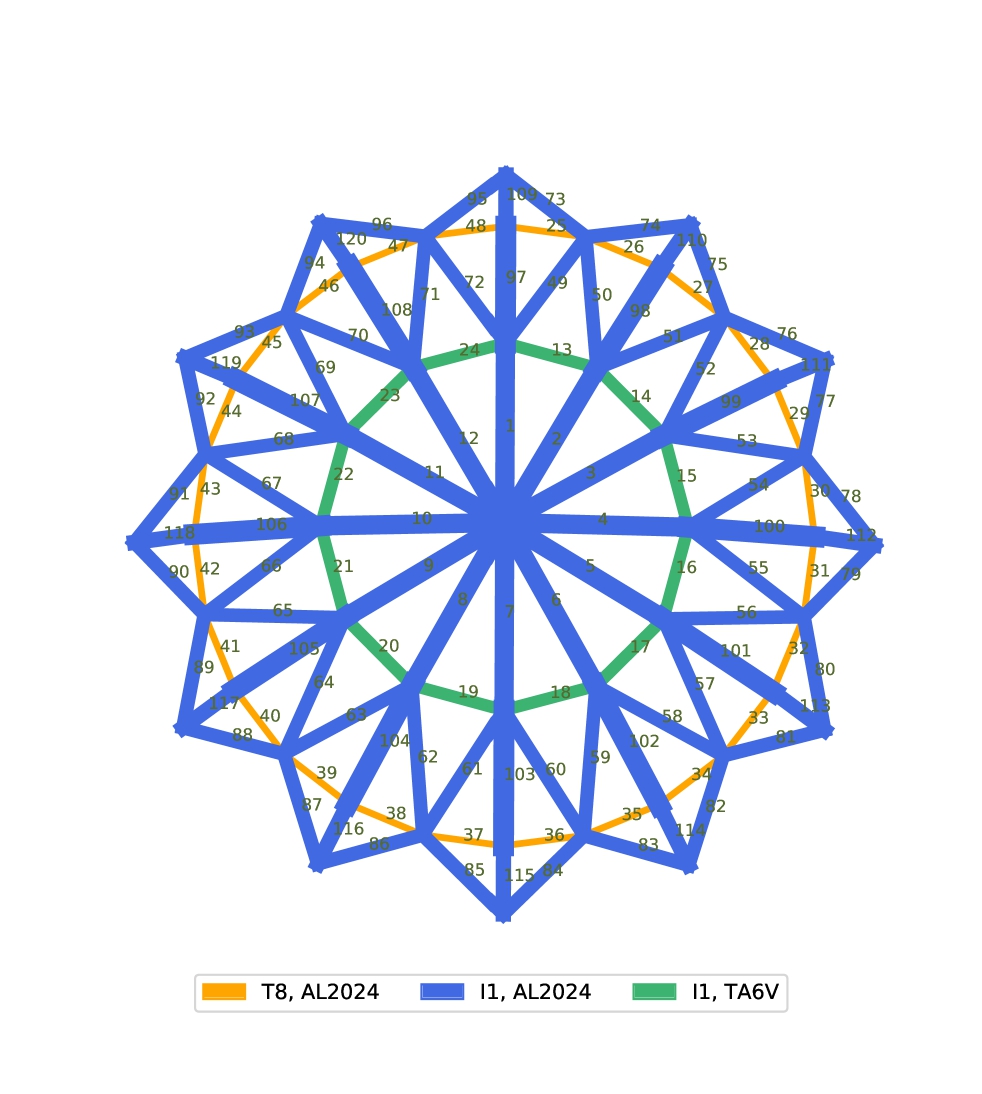}
\end{center}
\caption{Top view of the 120-bar truss mixed categorical-continuous optimization result.}
\label{fig:120bars_solution_top}
\end{figure}

\begin{figure}
\captionsetup{width=0.5\textwidth}
%\hspace*{1.5cm}
\begin{center}
\includegraphics[width=0.5\textwidth]{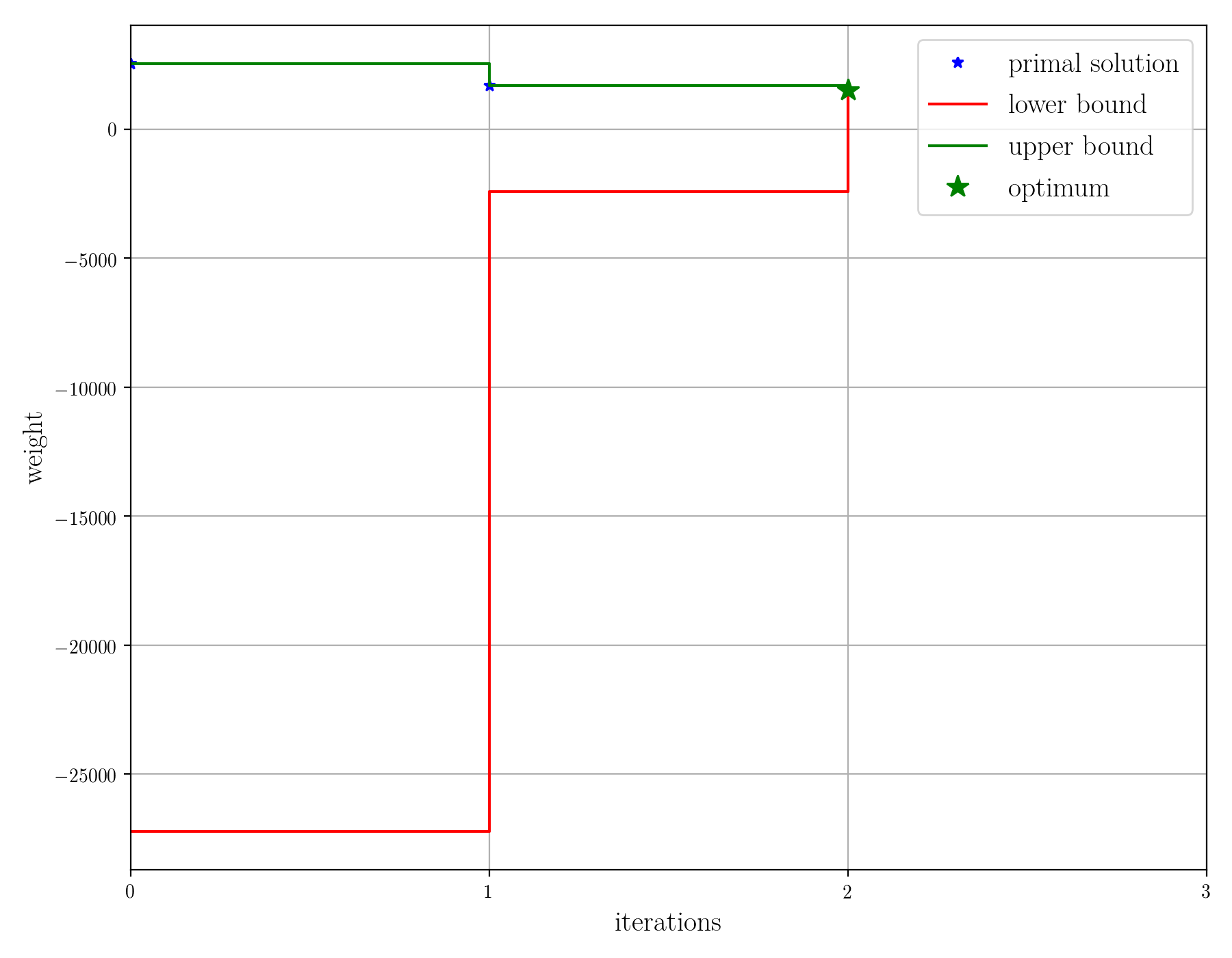}
\end{center}
\caption{History of the convergence of $\eta^{(k)}$ (i.e., the lower bound) and  $U^{(k)}$ (i.e., the upper bound) during the application of the \textsf{Bi-level OA} method to solve the 120-bar truss problem.% instance with 90 possible choices per structural element, so that $\vec B \in \mathcal{C}^{120,90}$.
}
\label{fig:120b_90c_oa}
\end{figure}

For this test problem, all the previously tested approaches (i.e., \textsf{Genetic}, \textsf{h-B\&B} and \textsf{Bi-level}) were unable to provide an optimum in a reasonable time (we could not converge to a competitive solution in 24 hours).
Only the \textsf{Bi-level OA} solver was able to converge to a competitive solution in approximately two hours; the optimal weight returned by \textsf{Bi-level OA} is $1506 ~kg$. Both the maximum displacement and the Euler buckling constraints interferes at the optimal solution. The Euler buckling constraints were active for all the structural elements, excepted the elements on the outer ring. Similarly to the previous test case, the local buckling constraints were not active at the solution.
The optimal truss is pictured Fig. \ref{fig:120bars_solution_top}.
The categorical and continuous solution is provided in Table \ref{tab:120bars_solution_90c} in Appendix \ref{apdx:120b_case_inputs}. We observe that only 3 choices have been selected over a total of 90 . The material ``TA6V'' has been selected for members 13 to 24 (inner circle, in green), and ``AL2024'' for the rest of the structure. The profile ``T8'' has been selected for members 25 to 48 (outer circle, in orange), while ``I1'' is selected for the rest of the structure.
The convergence history of $\eta^{(k)}$ (i.e., the lower bound) and  $U^{(k)}$ (i.e., the upper bound) is depicted Fig. \ref{fig:120b_90c_oa}. One can see that
the optimization process is converging within 3 iterations. 
This means that it required to solve only 3 NLP (primal problems), within a total of 35339 calls to FEM.

% %%%%%%
% CONLUSION

%%%%%
\section{Conclusion}\label{sec:conclusion_oa}

In this paper, we solved a mixed categorical-continuous structural optimization problem with categorical variables i.e. non-relaxable and non-ordered. 
The proposed algorithm used a bi-level decomposition of (\ref{pb:OverallPbBinary}), and solved a  sequence of master and slave problems. 
The resulting algorithm, named \textsf{Bi-level OA}, relied on the theory of the OA algorithm \citep{Fletcher1994,Bonami2008,Grossmann2009} where the derivatives are estimated using a post-optimal sensitivity analysis \citep{Fiacco1976}. 
Under a convexity assumption, we were able to guarantee the convergence of our proposed approach.

The numerical tests showed that the proposed method is capable of handling large scale instances of the mixed categorical-continuous problem. The scalability in terms of computational cost, has been tested with respect to the number of structural elements and number of categorical choices per element. Our convergence proof of the \textsf{Bi-level OA} relies on the convexity assumption with respect to the design variables, such assumption cannot be verified in the context of structural optimization problems.  A further work could consist in studying convergence of the proposed strategy when non-convex cases occur.

\section{Replication of results}
All the results in this paper are obtained using homemade Python code relying on the GEMS library, that will be released under open source license in 2021.
All the necessary data and equations are available in this paper to reproduce the results.
The geometries are depicted on Fig. \ref{fig:xbars_truss_oa} and Fig. \ref{fig:120bars_truss_oa}, material data is provided in Table \ref{tab:materialDefinition}, and the solution of the 120-bar truss is given in Table \ref{tab:120bars_solution_90c}.
The NLP and MINLP optimizations are performed thanks to the NLOPT \citep{Johnson} package and Coin-or Branch and Cut wrapped into the Google ortools suite \citep{ortools}, respectively.
Both libraries are under open source license.

\section*{Compliance with ethical standards}

\par\smallskip

\noindent\textbf{Conflict of interest}
 The authors declare that they have no conflict of interest.

\section*{Acknowledgments}
This work is part of the MDA-MDO project of the French Institute of Technology IRT Saint Exupery. We would lke to thank D. Bettebghor for his help and advices.  We are also very thankful for two anonymous reviewers; their insightful comments improved significantly the content of the paper.
We wish to
acknowledge the PIA framework (CGI, ANR) and the project industrial members for their support, financial backing
and/or own knowledge: Airbus, Altran Technologies, Capgemini DEMS France and CERFACS. The support of F. Gallard (IRT Saint Exupery) for his scientific and technical assistance on multilevel schemes implementation is gratefully acknowledged. 

\bibliographystyle{bibliosmo} % spphys spmpsci spmpsci_unsrt spbasic spbasic_updated plainnat plainnat_smo
% with natlib : unsrtnat
\bibliography{biblio}

\begin{appendices}

\section{On the computation of $\frac{d\Psi}{d\vec B}\Big|_{\vec B^{(k)}}$ using post-optimal sensitivities }\label{apdx:postop_sensitivities}

Gradient estimation, using  post-optimal sensitivities, was introduced by \citep{Fiacco1976} using penalty approach. In the context of our structural optimization problem, the estimation of the gradient can be derived as follows. The Lagrangian of the problem (\ref{pb:binary_parameterized_continuous_opt}) is given by
\begin{align*}
    \mathcal{L}(\vec a, \vec B) := \widetilde w(\vec a, \vec B) + \vec \lambda^{\top}_{\vec s} \widetilde{\vec s}(\vec a, \vec B) + \vec \lambda^{\top}_{\vec \delta} \widetilde{\vec \delta}(\vec a, \vec B) \\ + \vec \lambda^{\top}_{\ubar{\vec a}} \left(\ubar{\vec a} - \vec{a}\right) + \vec \lambda^{\top}_{\bar{\vec a}} \left(\vec{a} - \bar{\vec a} \right),
\end{align*}
where $~\vec \lambda_{\vec s},
      ~\vec \lambda_{\vec \delta},
      ~\vec \lambda_{\ubar{\vec a}}, \mbox{and}
      ~\vec \lambda_{\bar{\vec a}}$ are
the Lagrange multipliers (column vectors) associated to the constraints $\widetilde{ \vec s}$, $\widetilde{ \vec \delta}$, $\ubar{\vec a}$, and $\bar{\vec a}$, respectively.

For a given $(k)$ iteration, let $\vec a^{(k)}$ be the solution of the of problem (\ref{pb:binary_parameterized_continuous_opt}) evaluated at $\vec B^{(k)}$ and define $\mathcal{A}^{(k)}_{\vec s}$, $\mathcal{A}^{(k)}_{\vec \delta}$, and $\mathcal{A}^{(k)}_{\ubar{\vec a}}$, $\mathcal{A}^{(k)}_{\bar{\vec a}}$ as the sets of active constraints, i.e., 
\begin{equation*}
\begin{aligned}
    & \mathcal{A}^{(k)}_{\vec s} = \left \{\forall i~|~\widetilde{\vec s}_{i}\left(\vec a^{(k)}\right) = 0 \right \}, ~
         \mathcal{A}^{(k)}_{\ubar{\vec{a}}} = \left \{\forall i~|~\vec a_i^{(k)} = \ubar{\vec{a}}_i \right \}, \\
    & \mathcal{A}^{(k)}_{\vec \delta} = \left \{\forall i~|~\widetilde{\vec \delta}_i\left(\vec a^{(k)}\right) = 0 \right \},~ 
     \mathcal{A}^{(k)}_{\bar{\vec{a}}} = \left \{\forall i~|~\vec a_i^{(k)} = \bar{\vec{a}}_i  \right \}.   
\end{aligned}
\end{equation*}
The active components of the constraints $\widetilde{\vec s}$ and $\widetilde{\vec \delta}$ will be noted by $\vec s^{(k)}_{\mathcal{A}^{(k)}_{\vec s}}$ and $\vec \delta^{(k)}_{\mathcal{A}^{(k)}_{\vec \delta}}$, respectively.
 The components of $\vec a^{(k)} $ whose indices belong to $\mathcal{A}^{(k)}_{\ubar{\vec{a}}}$ (resp. $\mathcal{A}^{(k)}_{\bar{\vec{a}}}$) will be noted by $\vec{a}^{(k)}_{\ubar{\mathcal{A}}}$   (resp. $\vec{a}^{(k)}_{\bar{\mathcal{A}}}$).
Similarly, the Lagrange multipliers  at the optimum will be denoted by 
$~\vec \lambda_{\vec s}^{(k)},
      ~\vec \lambda_{\vec \delta}^{(k)},
      ~\vec \lambda_{\ubar{\vec a}}^{(k)}$, and
      $\vec \lambda_{\bar{\vec a}}^{(k)}$. Again, the Lagrange multipliers  at the optimum related to the active constraints of $\widetilde{\vec s}$, $\widetilde{\vec \delta}$, and bounds constraints on $\vec a$ will be designated by noted 
      $~\vec \lambda^{(k)}_{\mathcal{A}^{(k)}_{\vec s}}$,$\vec \lambda^{(k)}_{\mathcal{A}^{(k)}_{\vec \delta}}$,$\vec \lambda_{\mathcal{A}^{(k)}_{\ubar a}}^{(k)}$, and 
      $\vec \lambda_{\mathcal{A}^{(k)}_{\bar a}}^{(k)}$, respectively.

%Let be $\vec I_{\mathcal{A}^{(k)}_{\ubar{\vec a}}} \in \mathcal{M}_{\vert\mathcal{A}^{(k)}_{\ubar{\vec a}}\vert, n}(\mathbb{R})$ and $\vec I_{\mathcal{A}^{(k)}_{\bar{\vec a}}} \in \mathcal{M}_{\vert\mathcal{A}^{(k)}_{\bar{\vec a}}\vert, n}(\mathbb{R})$ the gradients of the lower and upper bound constraints, respectively.
Assuming that the objective and the constraints functions of (\ref{pb:binary_parameterized_continuous_opt}) are continuously differentiable at $\vec a^{(k)}$ and that the gradients of active constraints at $\vec a^{(k)}$ are linearly independent.  Then, by using the Karush-Kuhn-Tucker (KKT) optimality conditions applied to (\ref{pb:binary_parameterized_continuous_opt}) at $\vec B^{(k)}$, one gets 
\begin{small}
\begin{eqnarray}\label{eq:lagr_multipliers}
%\begin{split}
        \dfrac{\partial \widetilde{w}}{\partial \vec a}\Bigg|_{\vec{z}^{(k)}} 
          +  \left[\vec \lambda^{(k)}_{\mathcal{A}^{(k)}_{\vec s}}\right]^{\top} \dfrac{\partial \widetilde{\vec s}_{\mathcal{A}^{(k)}_{\vec s}}}{\partial \vec a}\Bigg|_{\vec{z}^{(k)}} 
    +  \left[\vec \lambda^{(k)}_{\mathcal{A}^{(k)}_{\vec \delta}}\right]^{\top} \dfrac{\partial \widetilde{\vec \delta}_{\mathcal{A}^{(k)}_{\vec \delta}}}{\partial \vec a}\Bigg|_{\vec{z}^{(k)}} & & \\     
          - \left[\vec \lambda_{\mathcal{A}^{(k)}_{\ubar a}}^{(k)}\right]^{\top} \vec I_{\mathcal{A}^{(k)}_{\ubar{\vec a}}}
           + \left[\vec \lambda_{\mathcal{A}^{(k)}_{\bar a}}^{(k)} \right]^{\top} \vec I_{\mathcal{A}^{(k)}_{\bar{\vec a}}}&=&\vec{0}_{n},
          \nonumber
%\end{split} 
\end{eqnarray}
\end{small}
where $\vec{z}^{(k)}:=(\vec a^{(k)}, \vec B^{(k)})$ and the  notation $ g|_{\vec{z}} $ is used to denote the value of the function $g$ at the point $z$. The matrices $I_{\mathcal{A}^{(k)}_{\bar{\vec a}}} \in \mathbb{R}^{|\mathcal{A}^{(k)}_{\bar{\vec a}}| \times n}$ and $I_{\mathcal{A}^{(k)}_{\ubar{\vec a}}} \in \mathbb{R}^{|\mathcal{A}^{(k)}_{\ubar{\vec a}}| \times n}$ are such that $\forall j \in \llbracket 1, n\rrbracket$, one has
\begin{small}
\begin{eqnarray*}
   \forall i \in \mathcal{A}^{(k)}_{\ubar{\vec a}}, ~~\left(\vec I_{\mathcal{A}^{(k)}_{\ubar{\vec a}}} \right)_{ij} = \delta_{ij}
    &\mbox{and} &    \forall i \in \mathcal{A}^{(k)}_{\bar{\vec a}}, ~~
  \left(\vec I_{\mathcal{A}^{(k)}_{\bar{\vec a}}} \right)_{ij} = \delta_{ij}
\end{eqnarray*}
\end{small}
with $\delta_{ij}$ being the Kronecker symbol.

Consequently, once the problem (\ref{pb:binary_parameterized_continuous_opt}) is solved for a given choice of $\vec B^{(k)}$, the Lagrange multipliers corresponding to active constraints can be obtained by solving the linear system given by \eqref{eq:lagr_multipliers}. We note that, according to the KKT conditions, the computed values of the Lagrange multipliers have to be non-negative.

Now, under appropriate assumptions and by using~\cite[Theorem 2.1]{Fiacco1976}, one can deduce that the function $\Psi$ is continuously differentiable at $\vec B^{(k)}$. In fact, assuming that at each iteration $(k)$ of our optimization process, one has
\begin{itemize}
      \item[-] the functions $\widetilde{w}$, $\widetilde{\vec s}$, and $\widetilde{\vec \delta}$ are twice continuously differentiable w.r.t. $\vec a$.
      \item[-] $\dfrac{\partial \widetilde{w}}{\partial \vec a}$, $\dfrac{\partial \widetilde{\vec s}}{\partial \vec a}$, $\dfrac{\partial \widetilde{\vec \delta}}{\partial \vec a}$ are once continuously differentiable w.r.t. $\vec B$ in a neighborhood of $\vec{z}^{(k)}$,
      \item[-] the second order sufficient KKT conditions related with the problem (\ref{pb:binary_parameterized_continuous_opt})  hold at $\vec{z}^{(k)}$,
      \item[-] a strict complementary holds, i.e., in the example of the displacements constraints :
\begin{align*}%\tag{\textbf{SCSC}}
    \label{eq:SCSC}
    \vec (\lambda_{\vec \delta}^{(k)})_i= 0 \iff \vec \delta_i(\vec{z}^{(k)}) < 0 ~~  \forall i \in \mathcal{A}^{(k)}_{\vec \delta}.
\end{align*}
\end{itemize}

In this case, by using~\cite[Theorem 2.1]{Fiacco1976}, we conclude that the function $\Psi$ is continuously differentiable, and its derivative taken in $\vec B^{(k)}$ is given by
{\small
\begin{eqnarray}
%\begin{split}
 \dfrac{d\Psi}{d\vec B}\Bigg|_{\vec B^{(k)}} &= &
        \dfrac{\partial \widetilde{w}}{\partial \vec B}\Bigg|_{\vec{z}^{(k)}} 
          +  \left[\vec \lambda^{(k)}_{\mathcal{A}^{(k)}_{\vec s}}\right]^{\top} \dfrac{\partial \widetilde{\vec s}_{\mathcal{A}^{(k)}_{\vec s}}}{\partial \vec B}\Bigg|_{\vec{z}^{(k)}} \label{eq:post_opt_sensitivity_oa} \\
 & & \quad \quad \quad \quad   +      \left[\vec \lambda^{(k)}_{\mathcal{A}^{(k)}_{\vec \delta}}\right]^{\top} \dfrac{\partial \widetilde{\vec \delta}_{\mathcal{A}^{(k)}_{\vec \delta}}}{\partial \vec B}\Bigg|_{\vec{z}^{(k)}},
\nonumber
\end{eqnarray}
}where we used the fact that bound constraints on the areas do not depend on $\vec B$ in order to eliminate the terms related to the bounds in the right hand side of equation (\ref{eq:post_opt_sensitivity_oa}). 

In this section was have detailed the mathematical theory of the post-optimality sensitivity analysis as stated in \cite{Fiacco1976}, and applied to the problem (\ref{pb:binary_parameterized_continuous_opt}).
Hence, for a given $\vec a^{(k)}$, the gradient of the function $\Psi$ taken at $\vec B^{(k)}$ can be estimated in five main steps :
\begin{itemize}%[noitemsep,topsep=0pt]
  \item[-] Build the set of active constraints $\mathcal{A}^{(k)}_{{\vec s}}$, $\mathcal{A}^{(k)}_{{\vec \delta}}$, $\mathcal{A}^{(k)}_{\ubar{\vec a}}$, and $\mathcal{A}^{(k)}_{\bar{\vec a}}$.
  \item[-] Compute the gradients of the objective and active constraints w.r.t. $\vec a$ at the point $(\vec a^{(k)}, \vec B^{(k)})$.
  \item[-] Compute the Lagrange multipliers $~\vec \lambda^{(k)}_{\mathcal{A}^{(k)}_{\vec s}}$ and $\vec \lambda^{(k)}_{\mathcal{A}^{(k)}_{\vec \delta}}$ by solving the linear system (\ref{eq:lagr_multipliers}).
  \item[-] Compute the gradients of the objective and active constraints w.r.t. $\vec B$ at the point $(\vec a^{(k)}, \vec B^{(k)})$.
  \item[-] Compute the post-optimal sensitivity $\frac{d\Psi}{d\vec B}$ at $\vec B^{(k)}$ using equation (\ref{eq:post_opt_sensitivity_oa}).
\end{itemize}
For all our optimization test cases, we did not get any numerical issue while deriving the  post-optimality sensitivities. We thus believe that those assumptions are not strong on practical truss optimization.
\section{Derivatives of the weight and the constraint functions} \label{apdx:derivatives}

The derivatives of the weight function and the constraints (with respect to the areas $a$) are obtained by applying  the chain-rule theorem. Namely, the gradient analytic expression of the weight function \eqref{eq:weight_function} with respect to the areas $a$ is given by:
\begin{align*}
\dfrac{\partial \widetilde w}{\partial \vec a} = \left(\sum_{c=1}^{p} \rho(c) \ell_1 {\vec B}_{1c}, \ldots, \sum_{c=1}^{p} \rho(c) \ell_n {\vec B}_{nc} \right).
\end{align*}
The gradient of the constraints are obtained as follows
\begin{align*}
\dfrac{\partial \widetilde{\vec{s}}_{ij}}{\partial \vec a} & = \sum_{c=1}^p \vec{B}_{ic} \dfrac{\partial \vec{s}_{ij}}{\partial \vec a}(\vec{a}_i, c,
\vec{\Phi}_i(\vec{a}, \vec{B})) \dfrac{\partial \vec{\Phi}_i}{\partial \vec a}(\vec{a}, \vec{B}).
\end{align*}
The derivative of the internal axial force in each member of the structure $\dfrac{\partial \vec{\Phi}_i}{\partial \vec a}(\vec{a}, \vec{B})$ is given by 
{\small{
\begin{equation*}
    \dfrac{\partial \vec{\Phi}_i}{\partial \vec a}(\vec{a}, \vec{B}) = \widetilde{\bm K}^e_i(\vec{a}_i,\vec{B}_{i,:}) \vec T_i \bm u_i(\vec{a},\vec{B}) +
    \bm K^e_i(\vec{a}_i,\vec{B}_{i,:}) \vec T_i \dfrac{ \partial \bm u_i}{\partial \vec a}(\vec{a},\vec{B}).
\end{equation*}
}}
where $\widetilde{\bm K}^e_i(\vec{a}_i,\vec{B}_{i,:})=\dfrac{ \partial \bm K^e_i}{\partial \vec a}(\vec{a}_i,\vec{B}_{i,:}) $. The derivative of the displacements $\dfrac{ \partial \bm u_i}{\partial \vec a}(\vec{a},\vec{B})$ is obtained, after derivation of (\ref{eq:forces_equilibrium}), by  
$$
\dfrac{ \partial \bm u_i}{\partial \vec a}(\vec{a},\vec{B}) = - \left[\bm K^e_i(\vec{a}_i,\vec{B}_{i,:})\right]^{-1} \widetilde{\bm K}^e_i(\vec{a}_i,\vec{B}_{i,:}) u_i(\vec{a},\vec{B}).
$$
\section{The proposed Bi-level Algorithm}\label{apdx:bilevel_algorithm}
\begin{algorithm}
\caption{A Bi-level framework using OA  cuts.}\label{alg:bilevel_oa}
\begin{algorithmic}[1]
\State $\textbf{initialize~} \vec{B}^{(0)}, \epsilon>0, K^{(-1)}\gets \{\emptyset\}, U^{(-1)} \gets +\infty$. 
Set $\mbox{Feasible} \gets 1$ and $k \gets 0$.
\While{ Feasible = 1}
%(\ref{pb:relaxed_outer_approx_k}) is feasible} 
    \State Compute $\Psi(\vec B^{(k)})$, let $\vec a^{(k)}$ be the approximate solution to the problem (\ref{pb:binary_parameterized_continuous_opt}) evaluated at $\vec B^{(k)}$.
  \If {$\vec a^{(k)} \in \Omega(\vec B^{(k)})$ and  $\Psi(\vec B^{(k)}) < U^{(k)}$}  
  %    \State $\vec a^* \gets \vec a^{(k)}$
  %    \State $\vec B^* \gets \vec B^{(k)}$
      \State $U^{(k)} \gets \Psi(\vec B^{(k)})$
  \Else
      \State $U^{(k)} \gets U^{(k-1)}$
  \EndIf
  \State $K^{(k)} \gets K^{(k-1)}~\bigcup~\{ \vec B^{(k)} \}$
  \State Estimate $\frac{d\Psi}{d\vec B}$ at $\vec B^{(k)}$ as given in Section \ref{sssec:gradient_psi_oa}.
  \State Let $\vec B^{(k+1)}$ and $\eta^{(k+1)}$ be the approximate solution of (\ref{pb:relaxed_outer_approx_k}) 
  \If{{\small{ $\eta^{(k+1)}$ satisfies the constraints of (\ref{pb:relaxed_outer_approx_k})}}}
       \State $\mbox{Feasible} \gets 1$
  \Else
      \State $\mbox{Feasible} \gets 0$
  \EndIf
  \State $k \gets k+1$
\EndWhile
\State \textbf{return~}
$\vec{a}^{*}\gets \vec a^{(k-1)}$, $\vec{B}^*\gets \vec B^{(k-1)}$, and $\widetilde{w}^* \gets U^{(k-1)}$.
\end{algorithmic}
\end{algorithm}

\FloatBarrier

\section{Test cases data}\label{apdx:120b_case_inputs}
~~
\begin{table}[ht]%{0.45\linewidth}
\ra{1.3}
\setlength{\tabcolsep}{3pt}
\captionsetup{width=1\linewidth}
    \centering
    \begin{tabular}{r|ccc} %r|ccc
        & AL2139 & AL2024 & TA6V \\
        \hline %\midrule
        $\text{Density}~[kg/mm^3]$ & $2.8e^{-6}$ & $2.77e^{-6}$ & $4.43e^{-6}$\\
        $\text{Young modulus}~[\mathit{MPa}]$ & $7.1e^{4}$ & $7.4e^{4}$ & $11.0e^{4}$\\
        $\text{Poisson coefficient}~[-]$ & $0.3$ & $0.33$ & $0.33$\\
        $\text{Tension allowable}~[\mathit{MPa}]$ & $1.5e^{2}$ & $1.6e^{2}$ & $11.0e^{2}$ \\
        $\text{Compression allowable}~[\mathit{MPa}]$ & $2.0e^{2}$ & $2.1e^{2}$ & $8.6e^{2}$\\
    \end{tabular}
    \caption{Numerical details on materials attributes for the test cases.}
    \label{tab:materialDefinition}
\end{table}

\begin{table}[h!]%{0.45\linewidth}
%\tiny
%\ra{1}
%\setlength{\tabcolsep}{6pt}
\captionsetup{width=1\linewidth}
    \centering
    %\begin{minipage}{0.5\textwidth}
    \begin{tabular}{ccc} %r|ccc
        \toprule
        \textbf{elements} & catalog & $\vec a [mm^2]$ \\
        \hline %\midrule
        1  $\dots$ 12  & 62 & 1100 \\
        13 $\dots$ 24  & 32 & 695  \\
        25 $\dots$ 47  & 89 & 379  \\
        48 $\dots$ 72  & 62 & 773  \\
        73 $\dots$ 96  & 62 & 799  \\
        97 $\dots$ 108  & 62 & 799  \\
        109$\dots$ 120 & 62 & 1195
    \end{tabular}
    \caption{The obtained solution for the 120-bar truss problem.}
    \label{tab:120bars_solution_90c}
\end{table}

\begin{table}[h!]%{0.45\linewidth}
%\tiny
%\ra{1.3}
%\setlength{\tabcolsep}{6pt}
\captionsetup{width=1\linewidth}
    \centering
    \begin{tabular}{ccc} %r|ccc
        \toprule
        $c$ & \textbf{cross section} & \textbf{material} \\
        \hline %\midrule
         1 \dots 10   & $I_{1}$ \dots $I_{10}$ & AL2139  \\
        11 \dots 20   & $C_{1}$ \dots $C_{10}$ & AL2139  \\
        21 \dots 30   & $T_{1}$ \dots $T_{10}$ & AL2139  \\
        31 \dots 40   & $I_{1}$ \dots $I_{10}$ & TA6V    \\
        41 \dots 50   & $C_{1}$ \dots $C_{10}$ & TA6V    \\
        51 \dots 60   & $T_{1}$ \dots $T_{10}$ & TA6V    \\
        61 \dots 70   & $I_{1}$ \dots $I_{10}$ & AL2024  \\
        71 \dots 80   & $C_{1}$ \dots $C_{10}$ & AL2024  \\
        81 \dots 90   & $T_{1}$ \dots $T_{10}$ & AL2024  \\
    \end{tabular}
    \caption{A description of the categorical design space related to the 120-bar truss problem  ($c \in \{1, \dots, 90\}$).}
    \label{tab:catalogs}
\end{table}

\begin{table}[h!]%{0.45\linewidth}
%\ra{1.3}
%\setlength{\tabcolsep}{3pt}
\captionsetup{width=1\linewidth}
    \centering
    \begin{tabular}{r|ccc} %r|ccc
        & $\vec x_0[1]~[mm]$ & $\vec x_0[2]~[mm]$ & $\vec x_0[3]~[mm]$ \\
        \toprule %\hline %\midrule
        $I_1, C_1, T_1 $         & 5  &  50  & 40 \\
        $I_2 , C_2 , T_2 $       & 10 &  110 & 40 \\
        $I_3 , C_3 , T_3 $       & 10 &  90  & 40 \\ 
        $I_4 , C_4 , T_4 $       & 10 &  100 & 40 \\
        $I_5 , C_5 , T_5 $       & 5  &  100 & 40 \\
        $I_6 , C_6 , T_6 $       & 10 &  60  & 40 \\ 
        $I_7 , C_7 , T_7 $       & 15 &  100 & 40 \\
        $I_8 , C_8 , T_8 $       & 10 &  70  & 35 \\ 
        $I_9 , C_9 , T_9 $       & 10 &  80  & 40 \\ 
        $I_{10}, C_{10}, T_{10}$ & 10 &  90  & 45 \\ 
    \end{tabular}
    \caption{Definition of the profiles $I$, $C$ and $T$ reference detailed geometry of the 120-bar truss problem.}
    \label{tab:profilesDefinition}
\end{table}
\end{appendices}

\end{document}